\documentclass{aa}  

\usepackage{graphicx}
\usepackage{txfonts}
\usepackage[breaklinks=true]{hyperref}
\hypersetup{
    bookmarksopen=false,
    bookmarksnumbered=false,
    unicode=true,           
    pdftoolbar=true,        
    pdfmenubar=true,        
    pdffitwindow=false,     
    pdfstartview={FitH},    
    pdfnewwindow=true,      
    colorlinks=true,       
    linkcolor=red,          
    citecolor=magenta,        
    filecolor=green,      
    urlcolor=blue           
}

\usepackage{amssymb}
\usepackage{pifont}
\usepackage{todonotes}
\newcounter{todocounter}

\usepackage{amsmath}
\usepackage{multirow}

\begin{document}

   \title{Determining stellar atmospheric parameters and chemical abundances of FGK stars with iSpec
       \thanks{The code is available via \url{http://www.blancocuaresma.com/s/} } 
   }

   \author{ S. Blanco-Cuaresma \inst{1, 2}
            \and C. Soubiran \inst{1, 2}
            \and U. Heiter \inst{3} 
            \and P. Jofr\'e \inst{1, 2, 4}
          }
   \offprints{S. Blanco-Cuaresma, \email{blanco@obs.u-bordeaux1.fr}}

   \institute{Univ. Bordeaux, LAB, UMR 5804, F-33270, Floirac, France.
         \and
            CNRS, LAB, UMR 5804, F-33270, Floirac, France
         \and
            Department of Physics and Astronomy,  Uppsala University, Box 516, 75120 Uppsala, Sweden
         \and
            Institute of Astronomy, University of Cambridge, Madingley Road, Cambridge CB3 0HA, U.K.
            }


 
  \abstract
    {An increasing number of high-resolution stellar spectra is available today thanks to many past and ongoing extensive spectroscopic surveys. Consequently, the scientific community needs automatic procedures to derive atmospheric parameters and individual element abundances.}
   {Based on the widely known SPECTRUM code by R. O. Gray, we developed an integrated spectroscopic software framework suitable for the determination of atmospheric parameters (i.e., effective temperature, surface gravity, metallicity) and individual chemical abundances. The code, named iSpec and freely distributed, is written mainly in Python and can be used on different platforms.}
   {iSpec can derive atmospheric parameters by using the synthetic spectral fitting technique and the equivalent width method. We validated the performance of both approaches by developing two different pipelines and analyzing the Gaia FGK benchmark stars spectral library. The analysis was complemented with several tests designed to assess other aspects, such as the interpolation of model atmospheres and the performance with lower quality spectra.}
   {We provide a code ready to perform automatic stellar spectral analysis. We successfully assessed the results obtained for FGK stars with high-resolution and high signal-to-noise spectra.}
   {}


   \maketitle
%


\section{Introduction}

Ongoing high-resolution spectroscopic surveys such as the Gaia ESO survey \citep[GES,][]{2012Msngr.147...25G, 2013Msngr.154...47R} and the future HERMES/GALAH \citep{2010gama.conf..319F} provide an enormous amount of high-quality spectra, increasing the data already available in observatory's archives \citep{2004sf2a.conf..547M, 2006ASPC..351..690D, 2014PASP..126..469P}. This represents a unique opportunity to unravel the history of our Galaxy by studying the chemical signatures of large star samples.

This huge amount of spectroscopic data challenges us to develop automatic processes to perform the required analysis. Numerous automatic methods have been developed over the past years \citep[e.g.,][to name a few]{1996A&AS..118..595V, 1998A&A...338..151K, 2006MNRAS.370..141R, 2009A&A...501.1269K, 2010A&A...517A..57J, 2013ApJ...766...78M, 2013A&A...558A..38M} to treat the spectra and derive atmospheric parameters from large datasets.

The two most common approaches are the direct comparison of observed and synthetic spectra, and the use of equivalent width technique based on excitation equilibrium and ionization balance. Nevertheless, typically the implementations of those methods use different ingredients (e.g., atomic data, model atmospheres) and continuum normalization strategies, which hinders direct comparisons.

We developed a software framework, named iSpec, to easily treat spectral observations and derive atmospheric parameters by applying the two most popular strategies: the synthetic spectral fitting technique and the equivalent width method. The code provides a wide variety of options, which facilitates executing homogeneous analysis using the same continuum normalization strategy, model atmospheres, atomic information, and radiative transfer code (SPECTRUM from \citealt{1994AJ....107..742G}).

For the synthetic spectral fitting technique, iSpec compares an observed spectrum with synthetic ones generated on-the-fly, in a similar way as the tool Spectroscopy Made Easy (SME) does \citep{1996A&AS..118..595V}. A least-squares algorithm minimizes the difference between the synthetic and observed spectra. In each iteration, the algorithm varies one free parameter at a time and prognosticates in which direction it should move. Specific regions of the spectrum can be selected to minimize the computation time, focusing on the most relevant regions to better identify stars (i.e., wings of H-$\alpha$/MgI triplet, Fe I/II lines).

Regarding the equivalent widths method, iSpec fits Gaussian models to a given list of Fe I/II lines, and from their integrated area derives their respective equivalent width and thus their abundances. The algorithm for determining the atmospheric parameters is based on the same least-squares technique mentioned above, but the minimization criterion is linked to the assumption of excitation equilibrium and ionization balance, similar to GALA \citep{2013ApJ...766...78M} and FAMA \citep{2013A&A...558A..38M}.

iSpec was previously used to create a high-resolution spectral library of the Gaia FGK benchmark stars \citep{2014A&A...566A..98B}, which are a common set of calibration stars in different regions of the HR diagram and span a wide range in metallicity. The defining property of these stars is that we know their radius and bolometric flux, which allows us to obtain their effective temperature and surface gravity {\it fundamentally}, namely, independently of the spectra. 

\newcommand{\Teff}{$T_{\rm eff}$}
\newcommand{\logg}{$\log g$}
\newcommand{\Fbol}{$F_{\rm bol}$}
\newcommand{\Ang}{$\theta_{\rm LD}$}

For 34 FGK stars and M giants, angular diameters \Ang, bolometric fluxes \Fbol\, and parallaxes $\pi$ were extracted from the literature.
Stellar masses were determined from the comparison of effective temperature and luminosity to the output of stellar evolution models. We used two sets of models for most stars, provided by the Padova \citep{2008A&A...484..815B,2009A&A...508..355B} and Yonsei-Yale \citep{2003ApJS..144..259Y,2004ApJS..155..667D} groups, which cover a wide range of masses and metallicities.
With these input data, \Teff\ and \logg\ were derived from fundamental relations, independently of spectroscopy\footnote{$T_{\rm eff} = (F_{\rm bol}/\sigma)^{0.25} (0.5\,\theta_{\rm LD})^{-0.5}$ and $g = (GM)^2 (0.5\,\theta_{\rm LD}/\pi)^{-2}$, where $\sigma$ is the Stefan-Boltzmann constant and $G$ the Newtonian constant of gravitation}. The reference iron abundances were also derived by \citet{2014A&A...564A.133J}. 
A brief description of the Gaia FGK benchmark stars and their reference parameters is given in \citet{2013arXiv1312.2943J}. A detailed discussion of their fundamental \Teff\ and \logg\ values and comparison to \Teff\ and \logg\ values derived from high-resolution spectroscopy will be given in Heiter et al. (in prep.).

In this paper, we show how the library of the Gaia FGK benchmark stars can be used to assess and improve spectroscopic pipelines (in our case, based on iSpec) by comparing the derived atmospheric parameters and chemical abundances with the reference values.

The framework was designed to be flexible enough to be adapted to the needs of individual studies or extensive stellar surveys. iSpec can be used in automatic massive analysis through Python scripts, but it also includes a user-friendly visual interface that can easily interoperate with other astronomical applications such as TOPCAT\footnote{\url{http://www.star.bris.ac.uk/\string~mbt/topcat/}}, VOSpec\footnote{\url{http://www.sciops.esa.int/index.php?project=ESAVO\&page=vospec}} and splat\footnote{\url{http://star-www.dur.ac.uk/\string~pdraper/splat/splat.html}}, facilitating a indirect way to access the Virtual Observatory\footnote{\url{http://www.ivoa.net/}}

We describe the iSpec software framework in Sect.~\ref{s:framework}. The particularities of the pipelines developed for the current work, together with the tests and validations using the Gaia FGK benchmark star library are presented in Sect.~\ref{s:pipelines}. Additional general validations are reported in Sect.~\ref{s:analysis} and, finally, the conclusions can be found in Sect.~\ref{s:conclusions}.

\section{Spectroscopic software framework}\label{s:framework}

\subsection{Data treatment}\label{s:data_treatment}

The main functionalities for spectra treatment integrated into iSpec cover the following fundamental aspects:

\begin{enumerate}
    \item Continuum normalization: the continuum points of a spectrum are found by applying a median and maximum filter with different window sizes. The former smoothes out noisy and the later ignores deeper fluxes that belong to absorption lines (the continuum will be placed in slightly upper or lower locations depending on the values of those parameters). Afterwards, a polynomial or group of splines (to be chosen by the user) can be used to model the continuum, and finally the spectrum is normalized by dividing all the fluxes by the model.
    \item Resolution degradation: the spectral resolution can be degraded by convolving the fluxes with a Gaussian of a given full width at half maximum (km s$^{-1}$).
    \item Radial velocity: iSpec includes several observed and synthetic masks and templates for different spectral types that can be used to derive the radial velocity of the star by applying the cross-correlation technique \citep{2007AJ....134.1843A}.
    \item Telluric lines identification: telluric lines from Earth's atmosphere contaminate ground-observed spectra, and this can affect the parameter determination. Their position in the spectra can be determined by cross-correlating with a telluric mask built from a synthetic spectrum (from the TAPAS database, \citealt{2014A&A...564A..46B}).
    \item Re-sampling: spectra can be re-sampled by using linear (two points) or Bessel (four points) interpolation \citep{1998A&A...338..151K}.
    \item Equivalent width (EW) measurement: EWs are determined by fitting a Gaussian profile (a Voigt profile can also be chosen) in each absorption line and integrating its area.
\end{enumerate}

To analyze observed stellar spectra, it is commonly necessary to apply some of these operations to the reduced spectra. Nevertheless, it is possible to use third-party software (e.g., ARES from \citealt{2007A&A...469..783S}, DAOSPEC from \citealt{2008PASP..120.1332S}, or DOOp from \citealt{2014A&A...562A..10C}) for these steps in combination with iSpec for the subsequent analysis.

The Gaia FGK benchmark stars library was created integrally with iSpec and is a good example of the data treatment capabilities of the framework. An extensive description and exhaustive validation of these operations can be found in \cite{2014A&A...566A..98B}.

\subsection{Line selection}\label{s:line_selection}

To determinate atmospheric parameters and individual abundances, it is necessary to chose which absorption lines are going to be used. This selection will definitively affect the results, thus it is important to consider the level of reliability of the atomic data (i.e., oscillator strengths). iSpec provides the user with all the functionalities needed to perform a custom selection (e.g., line synthesis, theoretical equivalent width calculation, user interface for easy visual comparison). For instance, an effective approach to identify the lines with the best atomic data is described in \cite{2014A&A...561A..21S}.

In a second stage, iSpec also facilitates the quality assessment of the spectral regions that are going to be used in the analyses (see Sect.~\ref{s:obs_verification}). For instance, spectra might be affected by different levels of noise, cosmic rays, and telluric lines.

\subsection{Spectral synthesis and abundances from equivalent widths}\label{s:ingredients}

iSpec uses SPECTRUM to generate synthetic spectra and determinate abundances from equivalent widths. The framework includes all the basic ingredients needed for these purposes.

\subsubsection{Atomic line-lists}\label{s:atomic_linelists}

Several atomic line-lists are included in iSpec. They were previously transformed to the format that SPECTRUM requires:

\begin{enumerate}
    \item Central wavelength (\AA) of the absorption line.
    \item Species description formed by a combination of the atomic number and the ionization state (e.g., ``26.0'' and ``26.1'' refers to a neutral and ionized iron line, respectively).
    \item Lower and upper excitation energies ($\mathrm{cm}^{-1}$).
    \item Logarithm of the product of the statistical weight of the lower level and the oscillator strength for the transition (i.e., log(gf)).
    \item Fudge factor (parameter to adjust the line broadening due to poorly understood physical factors).
    \item Transition type indicating whether the $\sigma$ and $\alpha$ parameters used in the Anstee and O'Mara broadening theory are provided (coded as AO type) or the van der Waals broadening should be used (GA type). The individual broadening half-widths for Natural broadening and Stark broadening may also be specified.
\end{enumerate}

Some original line-lists provide the lower excitation energies in electron volts (eV), thus they were transformed to $\mathrm{cm}^{-1}$ by multiplying by a conversion factor (1 eV = $8065.544$ $\mathrm{cm}^{-1}$). When the upper excitation state was not provided, it was obtained by applying the following relation:
    \begin{equation}
        \mathrm{E}^{\mathrm{upper}} = \mathrm{E}_{\mathrm{lower}} + \frac{h~c}{\lambda} ~ 6.24150974 \times 10^{18},
    \end{equation} 
where the upper and lower excitation energies are in $\mathrm{cm}^{-1}$, $h$ is the Planck constant ($h = 6.62606957 \times 10^{-34} \mathrm{m}^2$ kg s$^{-1}$), $c$ the speed of light in vacuum ($c = 299792458.0$ m s$^{-1}$), $\lambda$ is the line wavelength position in meters and the final change in energy is converted to eV with a conversion factor (1J = $6.24150974\times10^{18}$ eV). Physical constants and conversion factors were taken from \cite{2012RvMP...84.1527M}. The fudge factor was disabled for all the lines (i.e., set to 1).

iSpec provides several ready-to-use atomic line-lists with wide wavelength coverage (from 300 to 1100 nm), such as the original SPECTRUM line-list, which contains atomic and molecular lines obtained mainly from the NIST Atomic Spectra Database \citep{2005MSAIS...8...96R} and Kurucz line-lists \citep{1995KurCD..23.....K}, and the default line-list extracted from the VALD database \citep{2011BaltA..20..503K} in February 2012.

\subsubsection{Abundances}\label{s:abundances}

iSpec provides a collection of ready-to-use solar abundances from \cite{1989GeCoA..53..197A}, \cite{1998SSRv...85..161G}, \cite{2005ASPC..336...25A}, \cite{2007SSRv..130..105G}, and \cite{2009ARA&A..47..481A}. SPECTRUM requires these abundances for the process of spectral synthesis, where the values will be scaled based on the target metallicity ([M/H]). Nevertheless, individual abundances can be fixed to a given unscaled value if required. iSpec's synthetic spectral fitting technique takes advantage of this functionality to derive individual chemical abundances from a given list of absorption lines. The related accuracy can be considerably improved when using a line-by-line differential approach \citep{2009A&A...508L..17R} because part of the biases in data treatment and errors in the atomic information will cancel out (specially if all the stars are of the same type).

\subsubsection{Pre-computed model atmospheres grid}\label{s:model_atmospheres}

iSpec incorporates different ATLAS\footnote{\url{http://kurucz.harvard.edu/grids.html}} \citep{2005MSAIS...8...14K} and MARCS\footnote{\url{http://marcs.astro.uu.se/}} \citep{2008A&A...486..951G} model atmospheres properly transformed to the format that SPECTRUM requires:
            
\begin{enumerate}
    \item $\int \rho\, dx$: mass depth (g cm$^{-2}$)
    \item $T$: temperature (K)
    \item $P_{\rm gas}$: gas pressure (dyn cm$^{-2}$)
    \item $n_e$: electron density (cm$^{-3}$)
    \item $\kappa_{\rm R}$: Rosseland mean absorption coefficient (cm$^{2}$ g$^{-1}$)
    \item $P_{\mathrm{rad}}$: radiation pressure (dyn cm$^{-2}$)
    \item $V_{\mathrm{mic}}$: microturbulent velocity (m s$^{-1}$)
\end{enumerate}

The original MARCS models do not provide the electron densities, thus they were derived with the following relation:
\begin{equation}
    n_{e} = \frac{P_{e}}{k ~ T},
\end{equation} 
where $P_{e}$ is the electron pressure (dyn cm$^{-2}$) present in the MARCS model atmospheres, and $k$ is the Boltzmann constant ($1.3806488 \times10^{-16}$ erg K$^{-1}$).

It is worth noting that the MARCS grid is formed by a combination of plane-parallel and spherical models (ATLAS is plane-parallel only). The first is adequate for modeling the atmosphere of dwarf stars, while the second is more appropriate for giant stars. However, SPECTRUM will interpret the spherical models as plane-parallel. The differences that may be introduced are not important for the F, G, and K giants, as shown by \cite{2006A&A...452.1039H}.

The included MARCS model atmospheres cover the 2500 to 8000 K range in effective temperature, 0.00 to 5.00 dex in surface gravity, and $-$5.00 to 1.00 dex in metallicity with standard abundance composition (Table \ref{tab:MARCS_composition}). The original ATLAS by \cite{2005MSAIS...8...14K} and subsequent versions computed by \cite{2011PASP..123..531K} and \cite{2012AJ....144..120M} for APOGEE \citep{2008AN....329.1018A}, cover the 4500 to 8750 K range in effective temperature, 0.00 to 5.00 dex in surface gravity and $-$5.00 to 1.00 dex in metallicity.

\begin{table}
    \caption{Standard abundance composition for pre-computed MARCS model atmospheres.}
    \label{tab:MARCS_composition}
    \begin{centering}
        \begin{center}
            \begin{tabular}{l|c|c|c|c}
            \textbf{[Fe/H]} & \textbf{[$\alpha$/Fe]} & \textbf{[C/Fe]} & \textbf{[N/Fe]} & \textbf{[O/Fe]}\\
            \hline
            +1.00 to 0.00 &  0.00 &   0.00  &  0.00  &  0.00\\
            $-$0.25 &  +0.10 &  0.00 &   0.00 &   +0.10\\
            $-$0.50 &  +0.20 &  0.00 &   0.00 &   +0.20\\
            $-$0.75 &  +0.30 &  0.00 &   0.00 &   +0.30\\
            $-$1.00 to $-$5.00 &  +0.40 &  0.00 &   0.00 &   +0.40\\
            \end{tabular}
        \end{center}

        \par
    \end{centering}
\end{table}

\subsubsection{Interpolation of model atmospheres}\label{s:model_atmospheres_interpolation}

\begin{figure}
    \begin{centering}
        \includegraphics[width=\linewidth, trim = 1mm 1mm 1mm 1mm, clip]{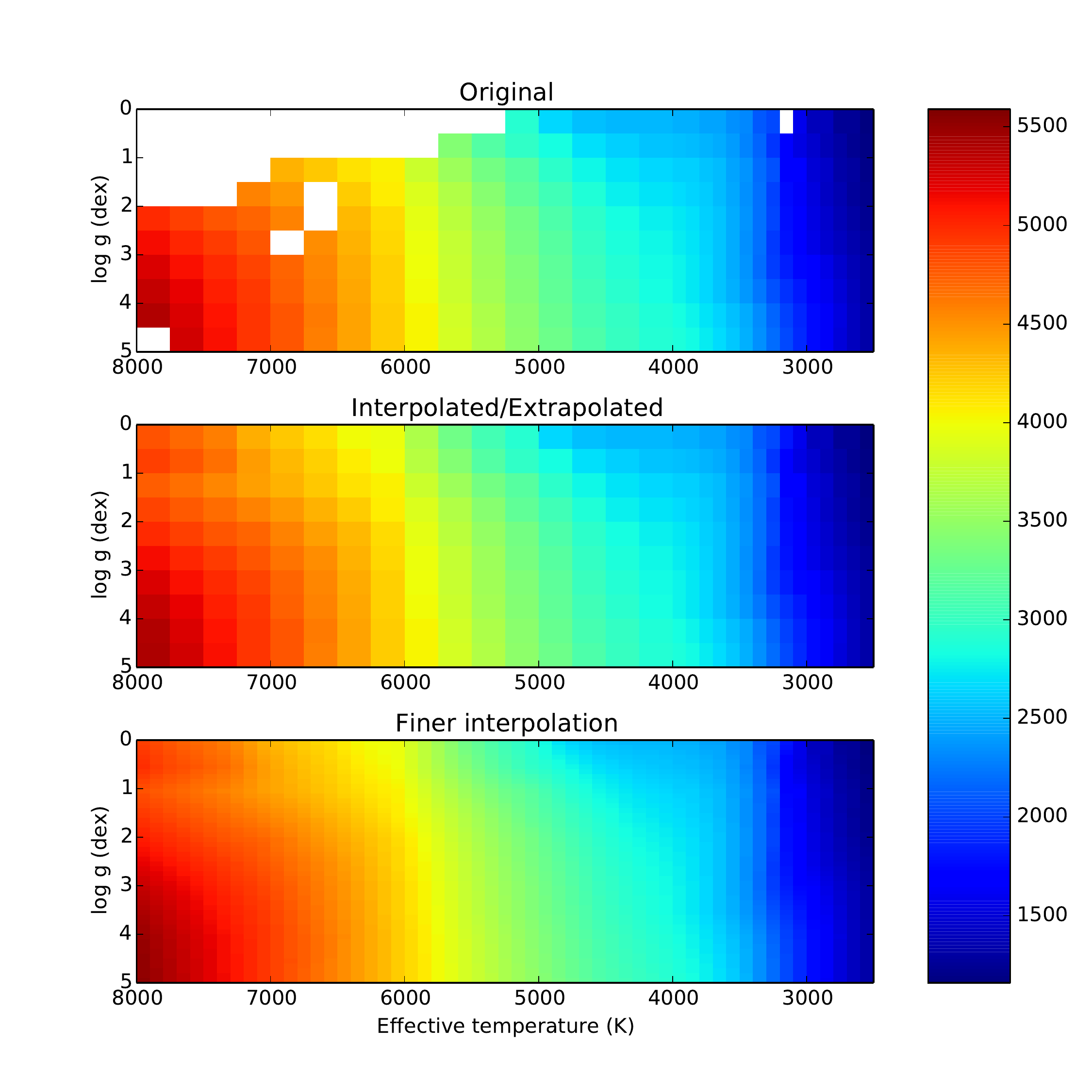}
        \par
    \end{centering}
    \caption{Original pre-computed MARCS model atmospheres for solar metallicity (upper), complete grid with interpolated and extrapolated atmospheres (middle), and homogeneously sampled grid (lower). The color scale represents the temperature for the first model atmosphere layer.}
    \label{fig:atm_layer0}
\end{figure}

The pre-computed model atmosphere grids presented in Sect. \ref{s:model_atmospheres} offer a reasonable coverage for the synthesis of typical FGK stars, but they do not provide a model for every single combination of effective temperature, surface gravity, and metallicity (e.g., white gaps in the upper plot from Fig. \ref{fig:atm_layer0} represent missing model atmospheres). On the other hand, the steps on effective temperature (typically $\sim$250 K), surface gravity ($\sim$ 0.5 dex) and metallicity ($\sim$ 0.50 dex) are not fine enough to optimally explore the parameter space.

We completed the model atmospheres grid as shown in the middle plot in Fig. \ref{fig:atm_layer0} by using interpolation and extrapolation. For the temperature-gravity combinations for which there is no pre-computed atmosphere, the procedure is as follows:

\begin{enumerate}
    \item First stage: linearly interpolate each value of each model atmosphere's layer from the existing pre-computed ones. In the best cases, two interpolated atmospheres are obtained and averaged:
        \begin{enumerate}
            \item \label{interp_fixed_logg} fixed gravity, and temperatures above and below the target parameters,
            \item \label{interp_fixed_teff} fixed temperature, and gravities above and below the target parameters.
        \end{enumerate}
    \item Second stage: linearly extrapolate. Again, in the best cases, two extrapolated atmospheres are obtained and combined with a weighted average that depends on their distance to the target atmosphere:
        \begin{enumerate}
            \item \label{averaged_extrap} fixed gravity, and two atmospheres with the closest temperature,
            \item \label{simple_extrap}   fixed temperature, and two atmospheres with the closest gravity.
         \end{enumerate}
        Extrapolated values were limited by the highest and lowest values found in all the real atmospheres to try to avoid unphysical results.
\end{enumerate}

All the quantities in the atmosphere (e.g., temperature, gas pressure, electron density) are re-sampled on
a common optical depth scale as described in \cite{2013MNRAS.430.3285M}. After the complete grid without gaps is constructed, it is easy to linearly interpolate any atmosphere that lies between any of the parameter combinations (i.e., effective temperature, surface gravity, and metallicity), as shown by the finer grid presented in the lower plot in Fig. \ref{fig:atm_layer0}. iSpec includes the complete interpolated version of the model atmosphere grids, only the in-between interpolation is made on-the-fly when requested by the user.

\subsection{Atmospheric parameter determination}\label{s:data_treatment}

iSpec is capable of determining atmospheric parameters and individual chemical abundances by using the synthetic spectral fitting technique and the equivalent width method.

In both cases, a $\chi^{2}$ minimization is performed by executing a nonlinear least-squares (Levenberg–Marquardt) fitting algorithm \citep{2009ASPC..411..251M}. The code starts from a given point in the parameter space and performs several iterations until convergence (i.e., the current/predicted $\chi^{2}$ is lower than a given threshold or the maximum number of iterations has been reached). In each iteration, the Jacobian is calculated via finite differences, linearizing the problem around the trial parameter set and changing each of the free parameters by a pre-established amount:
\begin{equation}
    p_{i} = p_{i}^{0} + \Delta{p_{i}},
    \label{eq:steps}
\end{equation}
where $p_{i}$ is a given free parameter, $p_{i}^{0}$ the current value, and $\Delta{p_{i}}$ the evaluation step.

\subsubsection{Initial parameters}\label{s:initial_params}

It is always recommended to provide initial parameters as close as possible to the expected final result, thus the computation time can be significantly reduced. Prior information such as photometry could be used for this purpose, but other approaches can be applied. For instance, iSpec provides the functionalities needed for computing a grid of synthetic spectra that can be used for deriving initial guesses via fast comparisons.

By default, iSpec does not include any pre-computed synthetic spectral grid since it strongly depends on the user requirements (i.e., wavelength ranges, atomic line-lists, model atmospheres). Therefore the user can employ the framework to build any custom grid and perform the initial parameter estimation.

\subsubsection{Synthetic spectral fitting}\label{s:synthetic}

The synthetic spectral fitting technique tries to minimize the difference between the observed and synthetic spectrum by directly comparing the whole observation or some delimited regions.

While exploring the parameter space, iSpec computes (via SPECTRUM) the synthetic spectra on-the-fly, which can be quite time-consuming depending on the extension of the chosen regions to be calculated and compared. Some codes prefer to approach this problem by executing the more time-consuming processes (i.e., pre-computing a huge grid of synthetic spectra) before starting the analysis, thus afterward the comparison time per star can be significantly faster. On the other hand, the on-the-fly approach is more flexible (e.g., it is very easy to adapt the analysis to different spectral resolutions).

The parameters that can be determined by using this method are the effective temperature, surface gravity, metallicity, microturbulence, macroturbulence, rotation, limb-darkening coefficient, and resolution. An efficient strategy is to let the first five parameters free and fix the rotation to 2 km s$^{-1}$ (since it degenerates with macroturbulence), limb-darkening coefficient to 0.6 and resolution to the one corresponding to the observation. After the atmospheric parameters are determined, individual chemical abundances can also be derived by the same method.

After several tests, we determined that the optimal step size for fastest convergence with the least-squares algorithm (Eq. \ref{eq:steps}) is 100 K for the effective temperature, 0.10 dex for surface gravity, 0.05 dex for metallicity, 0.50 km s$^{-1}$ for microturbulence, 2.0 km s$^{-1}$ for macroturbulence, 2.0 km s$^{-1}$ for rotation, 0.2 for the limb darkening coefficient, and 100 for the resolution.

\subsubsection{Equivalent width method}\label{s:ew}

The equivalent width method does not use all the information contained in the shape of the absorption-line profiles, but only their area. Therefore, broadening parameters such as rotation or macroturbulence are not considered.

iSpec derives (via SPECTRUM) individual abundances from the atomic data, the measured equivalent width, and a given effective temperature, surface gravity and microturbulence. The last three parameters are unknown and an initial guess is needed. iSpec gradually adjusts the atmospheric parameters by using neutral and ionized iron lines to enforce excitation equilibrium and ionization balance, which means that:

\begin{enumerate}
    \item Abundances as a function of the excitation potential should have no trends. If the trend is positive, the effective temperature is underestimated.
    \item Abundances as a function of the reduced equivalent width ($\mathrm{EWR} = \log_{10} \frac{\mathrm{EW}}{\lambda}$) should have no trends. If the trend is negative, the microturbulence is overestimated.
    \item The abundances of neutral iron (Fe 1) should be equal to the abundance of ionized iron (Fe 2). If the difference (Fe 1 - Fe 2) is positive, the surface gravity is underestimated.
\end{enumerate}

\begin{figure}
    \begin{centering}
        \includegraphics[width=\linewidth, trim = 1mm 1mm 1mm 1mm, clip]{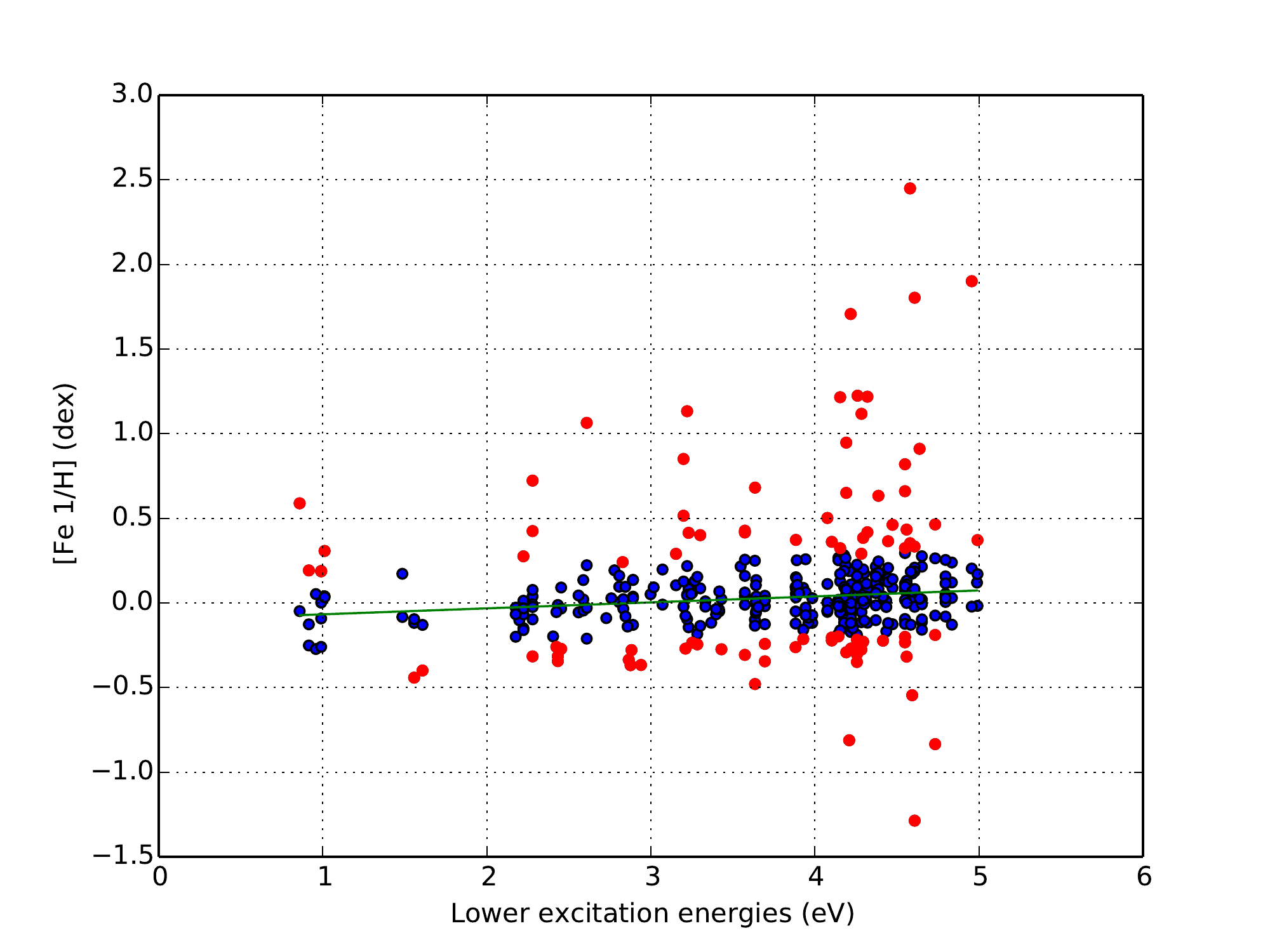}
        \par
    \end{centering}
    \caption{Iron abundances as a function of the excitation potential with a fitted linear model (green) and outlier values (red).}
    \label{fig:excitation_equilibrium_outliers}
\end{figure}

At the first iteration, iSpec identifies abundance outliers (see Fig. \ref{fig:excitation_equilibrium_outliers}) by robustly fitting a linear model using an M-estimator\footnote{``M'' for ``maximum-likelihood-type''}. If $r_{i}$ is the residual between the $i^{th}$ observation and its fitted value, a standard least-squares method would minimize $\sum_{i} r^{2}_{i}$ , which is strongly affected by outliers present in the data and distorts the estimation. The M-estimators try to reduce the effect of outliers by solving the following iterated re-weighted least-squares problem:

\begin{equation}
    \mathrm{min} \sum_{i} w \left( r^{k-1}_{i} \right) r^{2}_{i},
\end{equation}

where $k$ indicates the iteration number, and the weights $w \left( r^{k-1}_{i} \right)$ are recomputed after each iteration. After several tests, we determined that it is safe to discard abundances with assinged weights smaller than 0.90 (the weight scale ranges from zero to one). Robust regression estimators are shown to be more reliable than sigma clipping \citep{hekimoglu_outlier_2009}. 

After the initial abundance outliers are discarded, the least-squares algorithm minimizes three values: the two slopes of the linear models fitted as a function of the excitation potential and the reduced equivalent width and the difference in abundances from neutral and ionized iron lines.

The step size for the least square algorithm (Eq. \ref{eq:steps}) differs from the ones used in the synthetic spectral fitting technique. Based on different tests, we found that the optimal steps for fastest convergence are 500 K for the effective temperature, 0.50 dex for surface gravity, 0.05 dex for metallicity, and 0.50 km s$^{-1}$ for microturbulence.

\subsection{Error estimation}

The atmospheric parameter errors are calculated from the covariance matrix constructed by the nonlinear least-squares fitting algorithm. Nevertheless, they highly depend on the good estimation of the spectral flux errors. If they are underestimated, consequently, the atmospheric parameters errors will present unrealistically low values. The minimization process can be executed ignoring flux errors, but then the errors of the atmospheric parameters will be generally overestimated.

For the chemical abundance errors estimation, when possible (i.e., there is enough lines to measure), it is preferred to derive abundances for each line individually and consider the standard deviation as the internal error.

\section{Pipeline description and validation}\label{s:pipelines}

We developed two different pipelines based on the synthetic spectral fitting technique and the equivalent width method. In this section we describe some of their particularities and present results for different configurations (e.g., different model atmospheres).

\subsection{Line selection}\label{s:pipeline_line_selection}

\subsubsection{Atomic data verification}\label{s:atomic_data_verification}

For the current work, in addition to the atomic line lists included in iSpec by default, we also used the atomic data (without hyperfine structures and molecules) kindly provided by the GES line-list subworking group prior to publication (Heiter et al., in prep.). The line-list covers the optical range (i.e., 475 -- 685 nm) and provides a selection of medium and high-quality lines (based on the reliability of the oscillator strength and the blend level) for iron and other elements (e.g., Na, Mg, Al, Si, Ca, Sc, Ti, V, Cr, Mn, Co, Ni, Cu, Zn, Sr, Y, Zr, Ba, Nd, and Sm).

\subsubsection{Observation verification}\label{s:obs_verification}

The line selection is not only based on the quality of the atomic data, but also on the observed spectra to be analyzed. Using the iSpec framework, we fitted Gaussian profiles for all the selected absorption lines in each spectra. We automatically discarded lines that fall into one of these cases:

\begin{enumerate}
    \item Fitted Gaussian peak is too far away from the expected position (more than 0.0005 nm). Convection could produce shifts, but it is also possible that a strong close-by absorption line dominates the region and considerably blends the original targeted line. The analysis would require manual inspection, thus we reject those lines.
    \item Poor fits with extremely big root mean square difference (e.g., due to a cosmic ray)
    \item Potentially affected by telluric lines.
    \item Invalid fluxes (i.e., negative or nonexistent because of gaps in the observation).
\end{enumerate}

Additionally, only for the pipeline based on the equivalent width method and inspired by the GALA code, we filtered weak and strong lines based on their reduced equivalent widths. Weak lines are more sensitive to noise and errors in the continuum placement, while strong lines are usually significantly blended and may be severely affected by incorrect broadening parameters.

This verification process allowed us to adapt the analysis to the peculiarities of each observation, ensuring that only the best-quality regions were used for the final parameter determination.

\subsection{Atmospheric parameters determination}\label{s:analysis_pipelines}

The Gaia FGK benchmark stars library is a powerful tool for assessing the derived atmospheric parameters from different spectroscopic methods. We used the library to fine-tune our pipelines. The goal is to have results as similar as possible to the Gaia FGK benchmark stars reference values (accuracy) and with the lowest possible dispersion (precision).

\subsubsection{Initial parameters}\label{s:pipeline_initial_params}

To determinate the initial parameters (effective temperature, surface gravity, metallicity, micro/macroturbulence), we built a basic grid of synthetic spectra with iSpec for a selection of key parameters that allows us to easily separate giants from dwarfs and metal-rich from metal-poor stars (Table \ref{tab:synthetic_grid}). 

We compared the observed spectrum with all the spectra contained in the grid considering a selection of lines (see Sect. \ref{s:pipeline_line_selection}) and the wings of H-$\alpha$, H-$\beta$, and the magnesium triplet (around 515-520~nm). Finally, we adopted the parameters of the synthetic spectrum with the lowest $\chi^{2}$.

\begin{table}
    \caption{Synthetic spectral grid for determining the initial parameters. The rotation (v sin(i)) was fixed to $2$ km s$^{-1}$ and the micro/macroturbulence was calculated by using the same empirical relation as in \cite{2014A&A...566A..98B}.}
    \label{tab:synthetic_grid}
    \begin{centering}
        \begin{center}
            \begin{tabular}{l|c|c|c}
            \textbf{Class} & \textbf{T$_{\mathrm{eff}}$} & \textbf{log(g)} & \textbf{[M/H]}\\
            \hline
            Giant & 3500  &  1.00, 1.50  &  $-$2.00, $-$1.00, 0.00 \\
            Giant/Dwarf & 4500  &  1.50, 4.50  &  $-$2.00, $-$1.00, 0.00 \\
            Dwarf & 5500  &  4.50        &  $-$2.00, $-$1.00, 0.00 \\
            Dwarf & 6500  &  4.50        &  $-$2.00, $-$1.00, 0.00 \\
            \end{tabular}
        \end{center}

        \par
    \end{centering}
    \tablefoot{The micro/macroturbulence relation is based on GES UVES data release 1, the benchmark stars \citep{2014A&A...564A.133J}, and globular cluster data from external literature sources. }
\end{table}

\subsubsection{Synthetic spectral fitting method}\label{s:analysis_SSF}

In Table ~\ref{tab:results_synth_per_case}, we present the results for different configurations. In the interpretation, we consider a result as good when its differences are close to zero and its dispersion is low, but we prioritized lower dispersion over well centered values since it is easier to correct for a systematic shift than for a high dispersion. Among the different parameters, we also prioritized the surface gravity (even if that represents a slightly poorer result for the others) because it is the most difficult to derive when using only spectra.

\begin{enumerate}
    \item Line selection: the best results are obtained when combining lines from several elements (see Sect. \ref{s:pipeline_line_selection}) with the wings of the H-$\alpha$, H-$\beta$, and the magnesium triplet.
    \item Reduced equivalent width (EWR) filter: we evaluated the results obtained without a filter (unlimited) and two levels of restriction. The strong filter discards lines with an EWR lower than $-$5.8 and higher than $-$4.65, which means that lines with an equivalent width lower than 8 m\AA~and higher than 111 m\AA at 500 nm were discarded. The relaxed filter discards lines with EWR lower than $-$6.0 and higher than $-$4.3. At 500 nm, this equals discarding all the lines with an equivalent width lower than 5 m\AA~and higher than 250 m\AA. The results clearly do not show a better strategy, thus we chose the relaxed filter to match the same configuration as for the equivalent width pipeline (see Sect.~\ref{s:analysis_ew_method}).
    \item Atomic line list: the Gaia ESO Survey line-list (without hyper-fine structure and molecules) was compared with a line list extracted from VALD with the default options (2012) and with the SPECTRUM line-list, which includes molecules (see Sect.~\ref{s:atomic_linelists}). For the SPECTRUM line-list, we obtained poorer surface gravity and metallicities precisions. On the other hand, the GES and VALD line-lists are similar. Again, we chose the GES line-list to match the equivalent width pipeline configuration (see Sect.~\ref{s:analysis_ew_method}).
    \item Model atmospheres and solar abundances: ATLAS models, independent of the chosen solar abundance, are not as precise in all three atmospheric parameters, although they show a lower dispersion for stars with multiple spectra. We obtained better precisions with MARCS models when we combined them with the solar abundances from \cite{2007SSRv..130..105G}, thus we opted for these models.
    \item Initial atmospheric parameters: we performed one analysis starting systematically from the same point in the parameter space for all the stars (effective temperature of 5000 K, 2.5 dex in surface gravity, and solar metallicity) and one starting from an initial guess per spectrum (see Sect.~\ref{s:pipeline_initial_params}). The results are very similar, showing that the minimization process works well with independence of the starting point. On the other hand, it is preferred to always perform the initial guess because the computation time is considerably reduced (from an average of 50 minutes to 36 minutes per spectrum, see Sect.~\ref{s:iterations}).
    \item Resolution: we downgraded the resolution of the library to match the resolving power of Giraffe ESO/VLT (HR21 setup used in GES), which corresponds to 16200, and re-adjusted the continuum by renormalizing with a linear model. The results show that the pipeline could be used effectively for high and medium-resolution spectra.
\end{enumerate}

\begin{table*}[ht!]
    \caption{Average difference and standard deviation (left) between the synthetic spectral fitting technique and the reference values. Average dispersion (right) for stars with multiple observed spectra.}
    \label{tab:results_synth_per_case}
    \begin{center}
        \begin{tabular}{l|c c c c c c|c c c}
        \hline
         & \multicolumn{6}{c}{\textbf{Differences}} & \multicolumn{3}{|c}{\textbf{Dispersion}}\\
                      & \multicolumn{2}{c}{\textbf{$\Delta$T$_{\mathrm{eff}}$}} & 
                        \multicolumn{2}{c}{\textbf{$\Delta$log(g)}} &
                        \multicolumn{2}{c|}{\textbf{$\Delta$[M/H]}} & 
                        \textbf{$\Delta$T$_{\mathrm{eff}}$} & 
                        \textbf{$\Delta$log(g)} & 
                        \textbf{$\Delta$[M/H]} \\
        \textbf{Case} & $\mu$ & $\sigma$ & $\mu$ & $\sigma$ & $\mu$ & $\sigma$ & & & \\
        \hline
        All elements + wings    &   $-$24 &   124 &   $-$0.11   &   0.21    &   0.01    &   0.14    &   15  &   0.06    &   0.01    \\
        Only iron + wings   &   $-$2  &   110 &   $-$0.13   &   0.34    &   0.00    &   0.18    &   18  &   0.05    &   0.02    \\
        Only iron   &   $-$2  &   129 &   $-$0.29   &   0.35    &   $-$0.04   &   0.16    &   19  &   0.05    &   0.02    \\
        \hline                                                                          
        Relaxed EWR limit   &   $-$24 &   124 &   $-$0.11   &   0.21    &   0.01    &   0.14    &   15  &   0.06    &   0.01    \\
        Strong EWR limit    &   $-$14 &   122 &   $-$0.10   &   0.19    &   0.05    &   0.16    &   20  &   0.05    &   0.02    \\
        Unlimited EWR   &   $-$11 &   119 &   $-$0.07   &   0.21    &   0.02    &   0.14    &   16  &   0.04    &   0.01    \\
        \hline                                                                          
        GES &   $-$24 &   124 &   $-$0.11   &   0.21    &   0.01    &   0.14    &   15  &   0.06    &   0.01    \\
        VALD    &   4   &   119 &   $-$0.09   &   0.18    &   0.06    &   0.14    &   15  &   0.06    &   0.01    \\
        SPECTRUM    &   6   &   158 &   $-$0.17   &   0.25    &   0.07    &   0.16    &   19  &   0.05    &   0.02    \\
        \hline                                                                          
        MARCS/Greveese 2007 &   $-$24 &   124 &   $-$0.11   &   0.21    &   0.01    &   0.14    &   15  &   0.06    &   0.01    \\
        MARCS/Asplund 2005  &   $-$28 &   125 &   $-$0.17   &   0.25    &   0.01    &   0.14    &   13  &   0.05    &   0.02    \\
        ATLAS/Grevesse 2007 &   $-$47 &   244 &   $-$0.12   &   0.27    &   0.00    &   0.22    &   15  &   0.04    &   0.01    \\
        ATLAS/Asplund 2005  &   $-$50 &   243 &   $-$0.16   &   0.26    &   0.00    &   0.22    &   16  &   0.04    &   0.01    \\
        \hline                                                                          
        Estimate initial AP &   $-$24 &   124 &   $-$0.11   &   0.21    &   0.01    &   0.14    &   15  &   0.06    &   0.01    \\
        Fixed initial AP    &   $-$27 &   122 &   $-$0.12   &   0.22    &   0.02    &   0.15    &   12  &   0.05    &   0.01    \\
        \hline                                                                          
        R = 70000   &   $-$24 &   124 &   $-$0.11   &   0.21    &   0.01    &   0.14    &   15  &   0.06    &   0.01    \\
        R = 16200   &   $-$31 &   139 &   $-$0.15   &   0.20    &   0.06    &   0.22    &   18  &   0.06    &   0.03    \\
        R = 16200/re$-$norm   &   $-$58 &   137 &   $-$0.10   &   0.16    &   $-$0.07   &   0.14    &   17  &   0.06    &   0.02    \\
        \hline

        \end{tabular}
    \end{center}
\end{table*}

\begin{figure}
    \begin{centering}
        \includegraphics[width=\linewidth, trim = 1mm 1mm 1mm 1mm, clip]{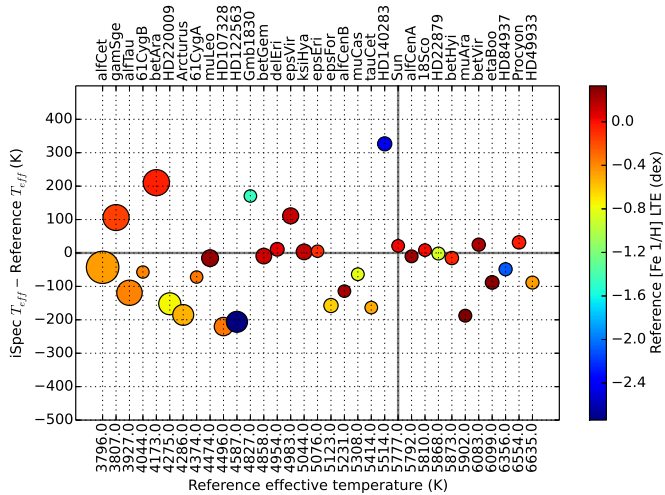}
        \par
    \end{centering}
    \caption{Differences in effective temperature between the reference (the Gaia FGK benchmark stars) and the derived value by iSpec (synthetic spectral fitting method). Stars are sorted by temperature; the color represents the metallicity, and larger symbols represent lower surface gravity.}
    \label{fig:bs_teff}
\end{figure}

\begin{figure}
    \begin{centering}
        \includegraphics[width=\linewidth, trim = 1mm 1mm 1mm 1mm, clip]{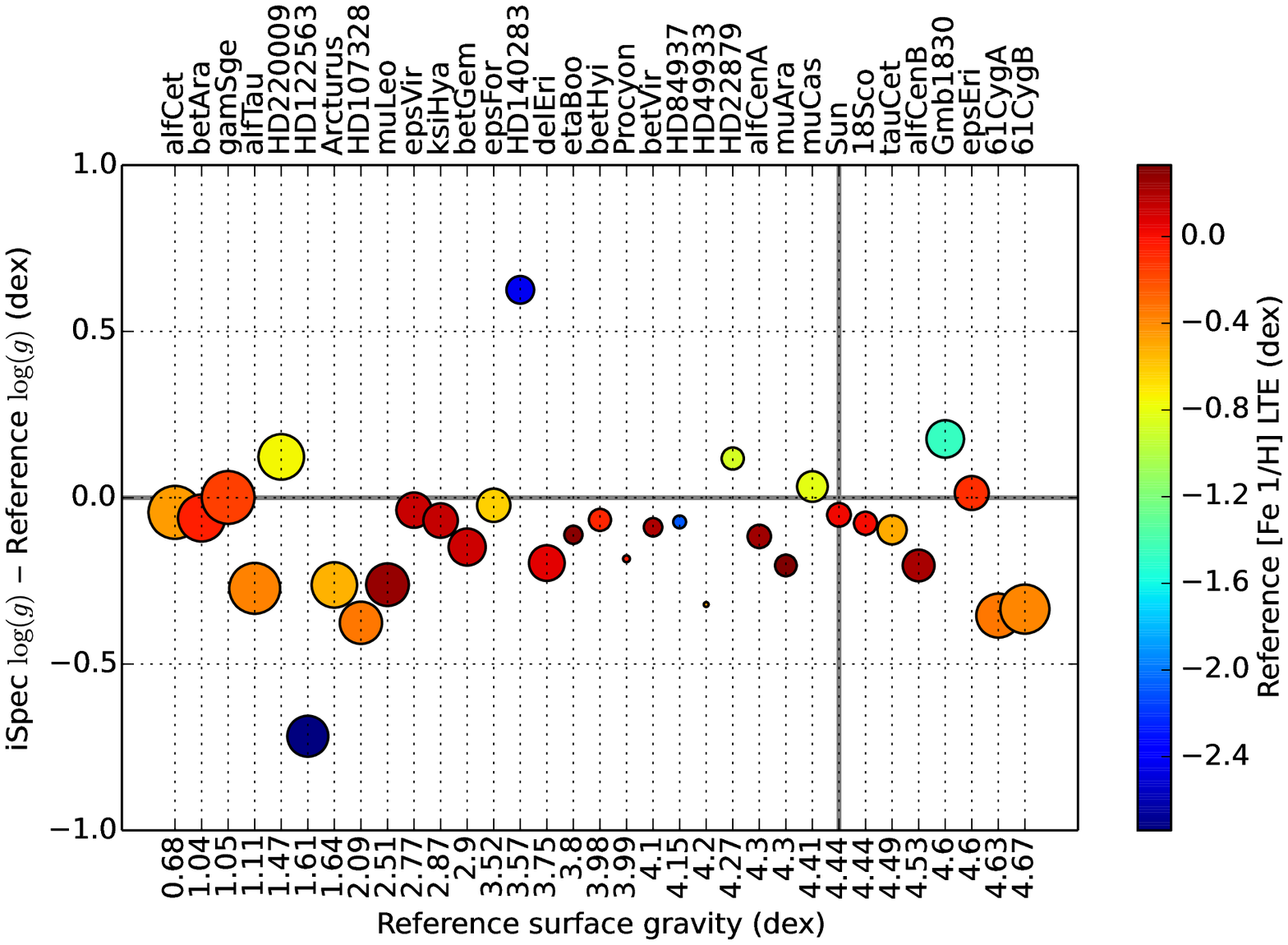}
        \par
    \end{centering}
    \caption{Differences in surface gravity between the reference (the Gaia FGK benchmark stars) and the derived value by iSpec (synthetic spectral fitting method). Stars are sorted by surface gravity; the color represents the metallicity, and larger symbols represent lower surface gravity.}
    \label{fig:bs_logg}
\end{figure}

The results per star with the best configuration can be found in the Table \ref{tab:detailed_results_for_stars_with_multiple_spectra} and Figs.~\ref{fig:bs_teff} and \ref{fig:bs_logg}. The average error (estimated by the least-square algorithm) is 19 K for effective temperature, 0.05 dex for surface gravity, and 0.02 dex for metallicity. These estimations have the same order of magnitude as the average dispersion obtained for the stars with several observed spectra: 15 K in effective temperature, 0.06 dex in surface gravity, and 0.01 dex in metallicity.


\begin{table*}[ht!]
    \caption{Difference between the parameters derived from the two methods and the reference values (neutral iron abundance is used as a proxy for metallicity). For stars with several observed spectra, the difference corresponds to the average, and the standard deviation is also reported.}
    \label{tab:detailed_results_for_stars_with_multiple_spectra}
    \begin{center}
        \tabcolsep=0.15cm
        \begin{tabular}{l|c c c|c c c c c c|c c c c c c}
        \hline
         & \multicolumn{3}{c|}{\textbf{Reference}} & \multicolumn{6}{c}{\textbf{Synthetic spectral fitting}} & \multicolumn{6}{|c}{\textbf{Equivalent width}}\\
                      & \textbf{T$_{\mathrm{eff}}$} & 
                        \textbf{log(g)} &
                        \textbf{[Fe 1/H]} & 
                        \multicolumn{2}{c}{\textbf{$\Delta$T$_{\mathrm{eff}}$}} & 
                        \multicolumn{2}{c}{\textbf{$\Delta$log(g)}} &
                        \multicolumn{2}{c|}{\textbf{$\Delta$[M/H]}} & 
                        \multicolumn{2}{c}{\textbf{$\Delta$T$_{\mathrm{eff}}$}} & 
                        \multicolumn{2}{c}{\textbf{$\Delta$log(g)}} & 
                        \multicolumn{2}{c}{\textbf{$\Delta$[M/H]}} \\
        \textbf{Star} & & & \tiny{$LTE$} & $\mu$ & $\sigma$ & $\mu$ & $\sigma$ & $\mu$ & $\sigma$ & $\mu$ & $\sigma$ & $\mu$ & $\sigma$ & $\mu$ & $\sigma$ \\
        \hline

18 Sco  &   5810    &   4.44    &   0.01    &   19  &   15  &   $-$0.06   &   0.03    &   0.03    &   0.01    &   $-$19 &   4   &   $-$0.04   &   0.03    &   0.03    &   0.01    \\
Arcturus    &   4286    &   1.64    &   $-$0.53   &   $-$179    &   8   &   $-$0.19   &   0.10    &   $-$0.11   &   0.02    &   $-$126    &   50  &   0.07    &   0.14    &   $-$0.13   &   0.04    \\
HD 107328   &   4496    &   2.09    &   $-$0.34   &   $-$218    &   3   &   $-$0.37   &   0.01    &   $-$0.18   &   0.01    &   $-$253    &   10  &   $-$0.28   &   0.01    &   $-$0.24   &   0.01    \\
HD 122563   &   4587    &   1.61    &   $-$2.74   &   $-$163    &   77  &   $-$0.61   &   0.32    &   0.00    &   0.05    &   434 &   30  &   0.61    &   0.13    &   0.57    &   0.04    \\
HD 140283   &   5514    &   3.57    &   $-$2.43   &   329 &   22  &   0.33    &   0.32    &   0.08    &   0.03    &   664 &   37  &   0.79    &   0.11    &   0.53    &   0.02    \\
HD 220009   &   4275    &   1.47    &   $-$0.75   &   $-$147    &   7   &   0.14    &   0.03    &   0.01    &   0.00    &   $-$10 &   41  &   0.56    &   0.13    &   0.04    &   0.03    \\
HD 22879    &   5868    &   4.27    &   $-$0.88   &   $-$1  &   1   &   0.13    &   0.02    &   0.16    &   0.00    &   $-$69 &   38  &   0.20    &   0.09    &   0.09    &   0.02    \\
HD 84937    &   6356    &   4.15    &   $-$2.09   &   $-$23 &   33  &   $-$0.05   &   0.06    &   0.17    &   0.04    &   291 &   76  &   0.55    &   0.09    &   0.46    &   0.08    \\
Procyon &   6554    &   3.99    &   $-$0.04   &   22  &   18  &   $-$0.20   &   0.02    &   $-$0.06   &   0.01    &   $-$35 &   75  &   0.24    &   0.17    &   0.00    &   0.01    \\
Sun &   5777    &   4.44    &   0.02    &   23  &   15  &   $-$0.06   &   0.02    &   $-$0.02   &   0.05    &   $-$4  &   25  &   $-$0.03   &   0.04    &   $-$0.01   &   0.03    \\
$\alpha$ Cen A  &   5792    &   4.30    &   0.24    &   $-$24 &   13  &   $-$0.13   &   0.01    &   $-$0.05   &   0.01    &   6   &   10  &   0.02    &   0.01    &   $-$0.04   &   0.00    \\
$\alpha$ Cet    &   3796    &   0.68    &   $-$0.45   &   $-$50 &   11  &   $-$0.01   &   0.04    &   0.22    &   0.01    &   317 &   36  &   1.02    &   0.25    &   0.24    &   0.02    \\
$\alpha$ Tau    &   3927    &   1.11    &   $-$0.37   &   $-$120    &   2   &   $-$0.20   &   0.10    &   0.10    &   0.00    &   347 &   75  &   0.70    &   0.71    &   0.23    &   0.21    \\
$\beta$ Hyi &   5873    &   3.98    &   $-$0.07   &   $-$25 &   9   &   $-$0.08   &   0.01    &   $-$0.06   &   0.01    &   26  &   9   &   0.13    &   0.01    &   0.00    &   0.01    \\
$\beta$ Vir &   6083    &   4.10    &   0.21    &   32  &   10  &   $-$0.08   &   0.01    &   $-$0.11   &   0.01    &   113 &   25  &   0.13    &   0.07    &   $-$0.02   &   0.01    \\
$\delta$ Eri    &   4954    &   3.75    &   0.06    &   16  &   9   &   $-$0.18   &   0.02    &   $-$0.01   &   0.01    &   284 &   30  &   0.20    &   0.12    &   0.08    &   0.04    \\
$\epsilon$ Eri  &   5076    &   4.60    &   $-$0.10   &   $-$13 &   16  &   $-$0.01   &   0.02    &   0.00    &   0.00    &   292 &   170 &   $-$0.29   &   0.18    &   0.02    &   0.00    \\
$\epsilon$ Vir  &   4983    &   2.77    &   0.13    &   111 &   0   &   $-$0.02   &   0.02    &   $-$0.05   &   0.00    &   262 &   21  &   0.69    &   0.07    &   0.04    &   0.02    \\
$\eta$ Boo  &   6099    &   3.80    &   0.30    &   $-$103    &   20  &   $-$0.12   &   0.02    &   $-$0.15   &   0.00    &   54  &   7   &   0.31    &   0.04    &   0.10    &   0.01    \\
$\mu$ Ara   &   5902    &   4.30    &   0.33    &   $-$196    &   11  &   $-$0.21   &   0.01    &   $-$0.09   &   0.00    &   $-$170    &   14  &   $-$0.10   &   0.03    &   $-$0.08   &   0.01    \\
$\tau$ Cet  &   5414    &   4.49    &   $-$0.50   &   $-$156    &   10  &   $-$0.08   &   0.03    &   0.05    &   0.01    &   $-$95 &   8   &   $-$0.28   &   0.03    &   0.04    &   0.02    \\
61 Cyg A    &   4374    &   4.63    &   $-$0.33   &   $-$72 &       &   $-$0.35   &       &   0.11    &       &   344 &       &   0.00    &       &   0.09    &       \\
61 Cyg B    &   4044    &   4.67    &   $-$0.38   &   $-$57 &       &   $-$0.33   &       &   0.17    &       &   135 &       &   $-$1.28   &       &   $-$0.36   &       \\
Gmb 1830    &   4827    &   4.60    &   $-$1.46   &   170 &       &   0.18    &       &   0.22    &       &   383 &       &   0.40    &       &   0.36    &       \\
HD 49933    &   6635    &   4.20    &   $-$0.46   &   $-$89 &       &   $-$0.32   &       &   $-$0.05   &       &   $-$39 &       &   0.24    &       &   0.14    &       \\
$\alpha$ Cen B  &   5231    &   4.53    &   0.22    &   $-$114    &       &   $-$0.20   &       &   $-$0.03   &       &   $-$64 &       &   $-$0.24   &       &   $-$0.06   &       \\
$\beta$ Ara &   4173    &   1.04    &   $-$0.05   &   210 &       &   $-$0.06   &       &   $-$0.06   &       &   51  &       &   $-$1.04   &       &   1.37    &       \\
$\beta$ Gem &   4858    &   2.90    &   0.12    &   $-$9  &       &   $-$0.15   &       &   $-$0.09   &       &   $-$57 &       &   $-$0.24   &       &   $-$0.14   &       \\
$\epsilon$ For  &   5123    &   3.52    &   $-$0.62   &   $-$157    &       &   $-$0.02   &       &   0.07    &       &   $-$43 &       &   0.04    &       &   0.09    &       \\
$\gamma$ Sge    &   3807    &   1.05    &   $-$0.16   &   106 &       &   0.00    &       &   0.04    &       &   233 &       &   1.08    &       &   0.36    &       \\
$\xi$ Hya   &   5044    &   2.87    &   0.14    &   3   &       &   $-$0.07   &       &   $-$0.11   &       &   156 &       &   0.50    &       &   $-$0.01   &       \\
$\mu$ Cas   &   5308    &   4.41    &   $-$0.82   &   $-$64 &       &   0.03    &       &   0.09    &       &   173 &       &   0.04    &       &   0.16    &       \\
$\mu$ Leo   &   4474    &   2.51    &   0.26    &   $-$16 &       &   $-$0.26   &       &   $-$0.03   &       &   157 &       &   $-$0.49   &       &   $-$0.26   &       \\
$\psi$ Phe  &   3472    &   0.51    &   $-$1.23   &   83  &       &   $-$0.35   &       &   0.92    &       &   941 &       &   0.30    &       &   0.65    &       \\

        \hline                                                  
        \end{tabular}
    \end{center}
\end{table*}

\subsubsection{Equivalent width method}\label{s:analysis_ew_method}

In the Table \ref{tab:results_ew_per_case}, we present the results for different configurations (the criteria for the interpretation are the same as in section \ref{s:analysis_SSF}):

\begin{enumerate}
    \item Iron line selection: a medium and high-quality selection of lines (495/286 neutral and 42/25 ionized iron lines) were tested, where the later is a subgroup of the former (see Sect.~\ref{s:atomic_data_verification}). The results show that the medium-quality group is preferred; a larger line sample seems to provide a higher statistical advantage for the equivalent width method.
    \item Reduced equivalent width filter: a filter based on different levels of the reduced equivalent width (as described in Sect.~\ref{s:analysis_SSF}) was applied. The highest effective temperature and surface gravity precision is achieved when using a relaxed limit, while strong limits show a better metallicity precision. We chose the former because we prefered to prioritize the surface gravity precision.
    \item Outliers filtering: the process of identifying outlier lines based on the derived abundance (see Sect.~\ref{s:ew}) was disabled to test its efficiency. The results clearly show that outlier filtering improves the accuracy and precision for all the parameters.
    \item Atomic line list: the results are better centered with the SPECTRUM line-list, but a lower dispersion is generally found for the GES line-list. Since we prioritize a low dispersion, we chose the GES line-list.
    \item Model atmospheres: the precision is very similar independently of the model atmosphere used, but MARCS produces slightly more accurate results.
    \item Initial atmospheric parameters: as described in Sect.~\ref{s:analysis_SSF}, we executed the analysis starting from a single point in the parameter space and compared this with starting from an initial guess (see Sect.~\ref{s:pipeline_initial_params}). The equivalent width method seems to be much more sensitive to the starting point, and it is highly recommended to start the analysis from a good initial guess.
    \item Resolution: absorption lines are more blended for lower spectral resolution. Consequently, the equivalent width method has more difficulties with low-resolution spectra, for which it over-estimates abundances and provides poorer results.
    \item Line profile: using Gaussian profiles to fit lines and derive equivalent widths seems to provide slightly better results than Voigt profiles.
\end{enumerate}

The results per star with the best configuration can be found in the Table \ref{tab:detailed_results_for_stars_with_multiple_spectra}. The average error (estimated by the least-squares algorithm)  is 67 K for effective temperature, 0.13 dex for surface gravity, and 0.09 dex for metallicity. These estimates have the same order of magnitude as the average dispersion obtained for the stars with several observed spectra: 38 K in effective temperature, 0.12 dex in surface gravity, and 0.03 dex in metallicity.

\begin{table*}[ht!]
    \caption{Average difference and standard deviation (left) between the equivalent width method and the reference values. Average dispersion (right) for stars with multiple observed spectra.}
    \label{tab:results_ew_per_case}
    \begin{center}
        \begin{tabular}{l|c c c c c c|c c c}
        \hline
         & \multicolumn{6}{c}{\textbf{Differences}} & \multicolumn{3}{|c}{\textbf{Dispersion}}\\
                      & \multicolumn{2}{c}{\textbf{$\Delta$T$_{\mathrm{eff}}$}} & 
                        \multicolumn{2}{c}{\textbf{$\Delta$log(g)}} &
                        \multicolumn{2}{c|}{\textbf{$\Delta$[M/H]}} & 
                        \textbf{$\Delta$T$_{\mathrm{eff}}$} & 
                        \textbf{$\Delta$log(g)} & 
                        \textbf{$\Delta$[M/H]} \\
        \textbf{Case} & $\mu$ & $\sigma$ & $\mu$ & $\sigma$ & $\mu$ & $\sigma$ & & & \\
        \hline

Medium-quality lines    &   135 &   241 &   0.20    &   0.44    &   0.12    &   0.27    &   38  &   0.12    &   0.03    \\
High-quality lines  &   129 &   311 &   0.14    &   0.47    &   0.09    &   0.30    &   39  &   0.09    &   0.06    \\
\hline                                                                          
Relaxed EWR limit   &   135 &   241 &   0.20    &   0.44    &   0.12    &   0.27    &   38  &   0.12    &   0.03    \\
Strong EWR limit    &   134 &   259 &   0.30    &   0.50    &   0.13    &   0.23    &   41  &   0.09    &   0.02    \\
Unlimited EWR   &   173 &   371 &   0.20    &   0.53    &   0.21    &   0.45    &   58  &   0.14    &   0.06    \\
\hline                                                                          
Outlier filter &   135 &   241 &   0.20    &   0.44    &   0.12    &   0.27    &   38  &   0.12    &   0.03    \\
With outliers   &   178 &   374 &   0.41    &   0.67    &   0.13    &   0.31    &   80  &   0.18    &   0.06    \\
\hline                                                                          
GES &   135 &   241 &   0.20    &   0.44    &   0.12    &   0.27    &   38  &   0.12    &   0.03    \\
VALD    &   118 &   313 &   0.20    &   0.58    &   0.19    &   0.24    &   57  &   0.12    &   0.06    \\
SPECTRUM    &   $-$1  &   326 &   $-$0.06   &   0.45    &   0.10    &   0.29    &   67  &   0.12    &   0.05    \\
\hline                                                                          
MARCS   &   135 &   241 &   0.20    &   0.44    &   0.12    &   0.27    &   38  &   0.12    &   0.03    \\
ATLAS   &   157 &   221 &   0.22    &   0.44    &   0.13    &   0.26    &   37  &   0.12    &   0.03    \\
\hline                                                                          
Estimate initial AP &   135 &   241 &   0.20    &   0.44    &   0.12    &   0.27    &   38  &   0.12    &   0.03    \\
Fixed initial AP    &   152 &   359 &   $-$0.23   &   0.81    &   0.08    &   0.26    &   49  &   0.18    &   0.04    \\
\hline                                                                          
R = 70000   &   135 &   241 &   0.20    &   0.44    &   0.12    &   0.27    &   38  &   0.12    &   0.03    \\
R = 16200   &   419 &   598 &   0.62    &   0.99    &   0.53    &   0.57    &   118 &   0.25    &   0.11    \\
R = 16200/re$-$norm   &   475 &   520 &   0.86    &   0.89    &   0.21    &   0.79    &   62  &   0.17    &   0.05    \\
\hline                                                                          
Gaussian profiles   &   135 &   241 &   0.20    &   0.44    &   0.12    &   0.27    &   38  &   0.12    &   0.03    \\
Voigt profiles  &   131 &   286 &   0.30    &   0.52    &   0.14    &   0.24    &   97  &   0.19    &   0.06    \\

        \hline                                                  
        \end{tabular}
    \end{center}
\end{table*}

\subsubsection{Method comparison}\label{s:method_comparison}

To our knowledge, this is the first time that these two methods were compared by covering such a wide range of parameters and using exactly the same normalization process, atomic data, model atmospheres, and radiative transfer code (i.e., SPECTRUM). The tests show that our pipeline based on the synthetic spectral fitting technique provides more accurate and precise atmospheric parameters (Table \ref{tab:results_per_method}), thus it is our preferred strategy when using iSpec.

On the other hand, it is worth noting that one of the advantages of other implementations of the equivalent width method (i.e., those based on the MOOG code \citealt{2012ascl.soft02009S}, such as GALA and FAMA) is their execution speed. SPECTRUM derives abundances from equivalent widths by completely synthesizing the lines, which makes the execution significantly slower (see Sect.~\ref{s:iterations}).

\begin{table}
    \caption{Average difference and standard deviation (left) between the derived parameters and the reference values. Average dispersion (right) for stars with multiple observed spectra.}
    \label{tab:results_per_method}
    \begin{center}
        \tabcolsep=0.08cm
        \begin{tabular}{l|c c c c c c|c c c}
        \hline
         & \multicolumn{6}{c}{\textbf{Differences}} & \multicolumn{3}{|c}{\textbf{Dispersion}}\\
                      & \multicolumn{2}{c}{\textbf{$\Delta$T$_{\mathrm{eff}}$}} & 
                        \multicolumn{2}{c}{\textbf{$\Delta$log(g)}} &
                        \multicolumn{2}{c|}{\textbf{$\Delta$[M/H]}} & 
                        \textbf{$\Delta$T$_{\mathrm{eff}}$} & 
                        \textbf{$\Delta$log(g)} & 
                        \textbf{$\Delta$[M/H]} \\
        \textbf{} & $\mu$ & $\sigma$ & $\mu$ & $\sigma$ & $\mu$ & $\sigma$ & & & \\
        \hline
        SSF   &   $-$24 &   124 &   $-$0.11   &   0.21    &   0.01    &   0.14    &   15  &   0.06    &   0.01    \\
        EW   &   135 &   241 &   0.20    &   0.44    &   0.12    &   0.27    &   38  &   0.12    &   0.03    \\
        \hline                                                  
        \end{tabular}
    \end{center}
    \tablefoot{SSF stands for synthetic spectral fitting technique and EW for equivalent width method. }
\end{table}

\subsection{Chemical abundances}

It is important to clarify that the metallicity parameter derived by iSpec is a global scale factor that is applied to all the elements (taking the solar abundance as the reference point) and it is not a direct measurement of the iron abundance (although it is a close approximation). Nevertheless, when the atmospheric parameters are known, iSpec can also determine individual chemical abundances for any element.

We limited our analysis to the determination of neutral and ionized iron since, up to now, it is the only element for which a reference value exists.

Our pipeline implements a line-by-line differential analysis (see \ref{s:abundances}) that follows these steps:

\begin{enumerate}
    \item Derive the solar absolute abundances for each of the selected lines (see Sect.~\ref{s:pipeline_line_selection}) using the reference atmospheric parameters. Seven out of the eight solar spectra included in the Gaia FGK benchmark stars library were used; we discarded one with a very low signal-to-noise ratio.
    \item Discard lines with a high dispersion among the seven solar spectra (interquartile range higher than 0.025).
    \item Derive the absolute abundances for each of the selected lines using the spectrum to be analyzed.
    \item Calculate the relative abundances by subtracting the solar absolute abundances to the absolute abundances of the spectrum being analyzed.
    \item Derive the final abundance for the spectrum by calculating the median value and consider the dispersion as the internal error.
\end{enumerate}

\begin{figure*}
    \begin{centering}
        \includegraphics[width=9cm, trim = 1mm 1mm 1mm 1mm, clip]{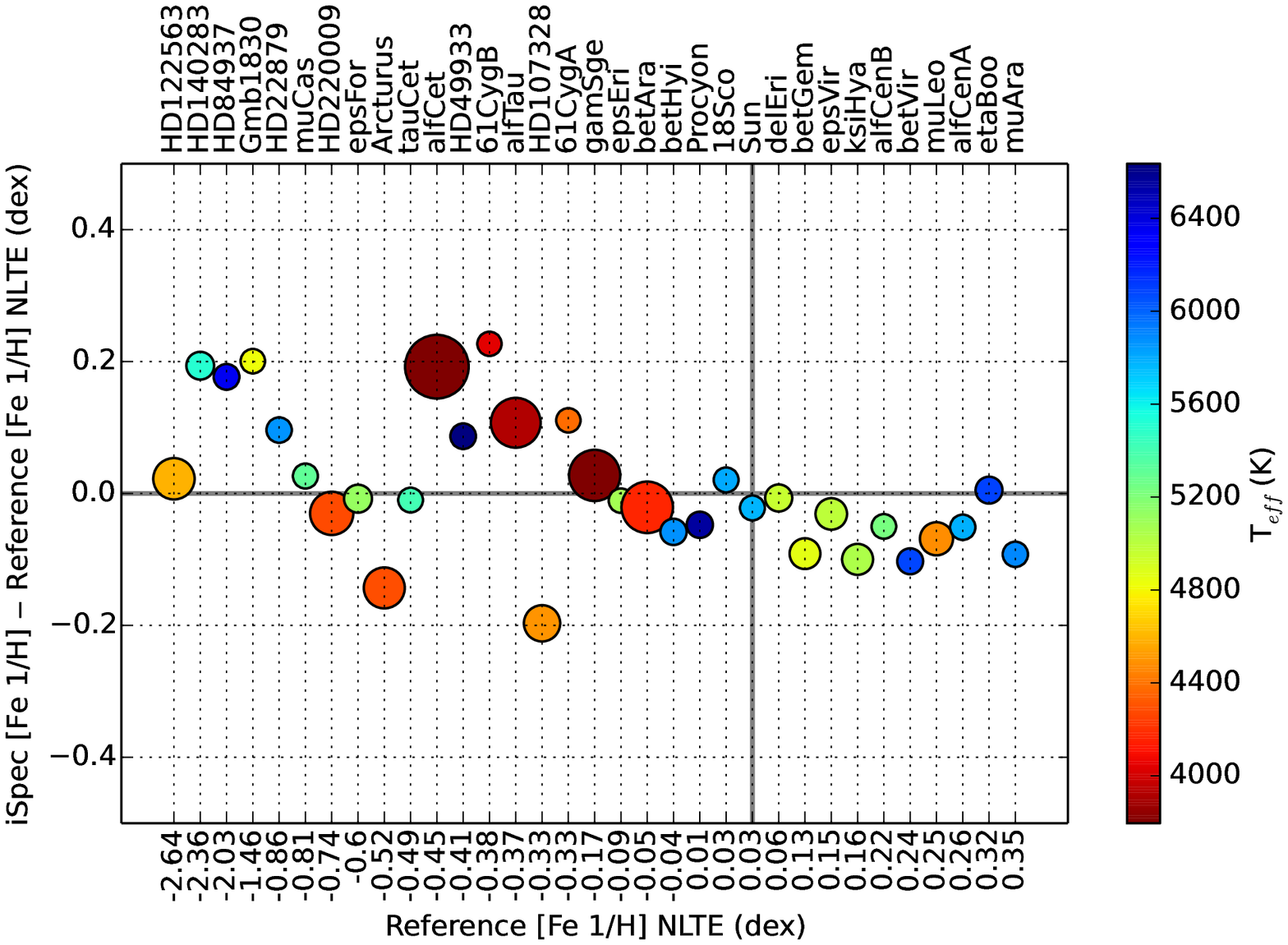}
        \includegraphics[width=9cm, trim = 1mm 1mm 1mm 1mm, clip]{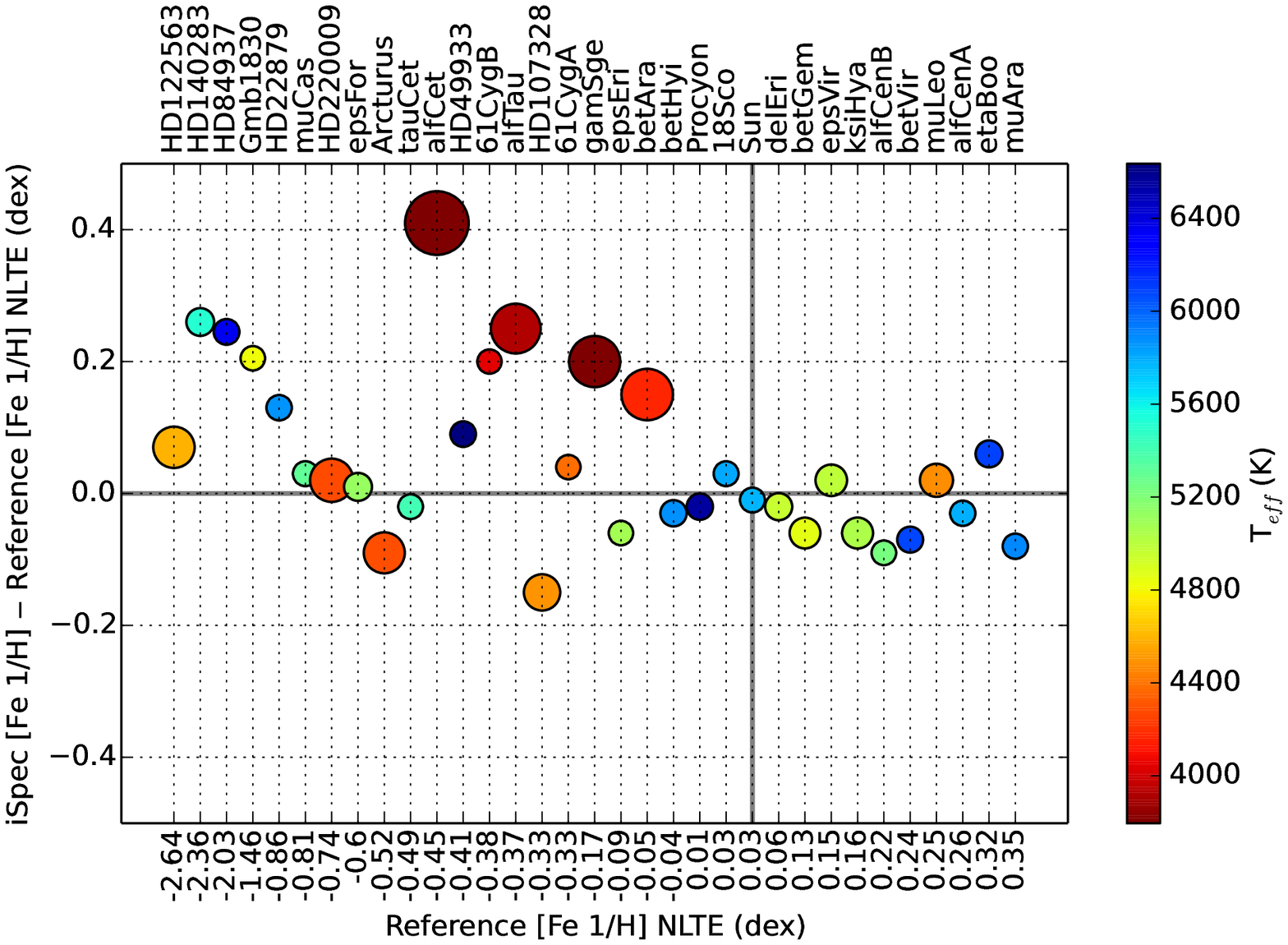}
        \par
        \includegraphics[width=9cm, trim = 1mm 1mm 1mm 1mm, clip]{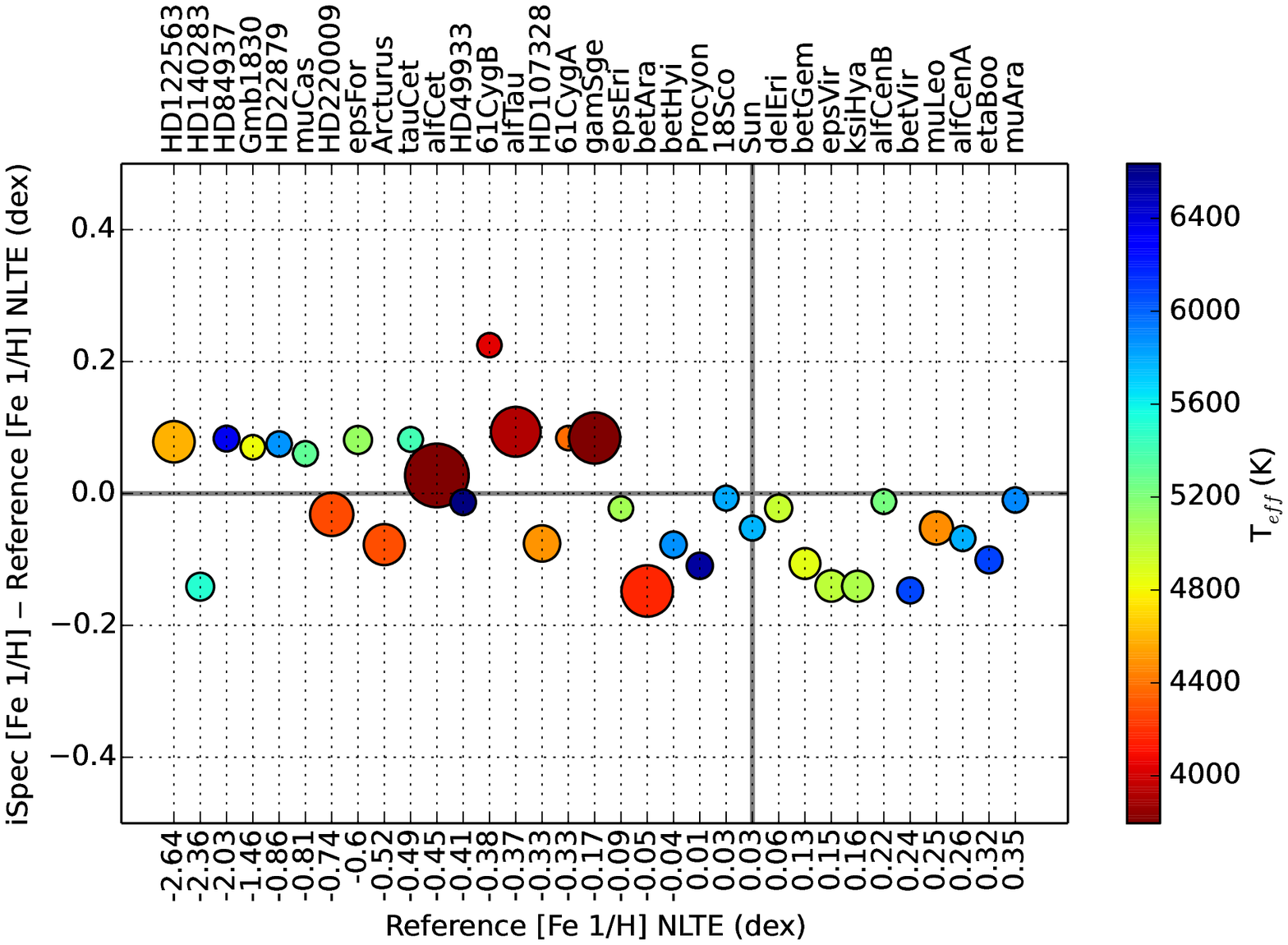}
        \includegraphics[width=9cm, trim = 1mm 1mm 1mm 1mm, clip]{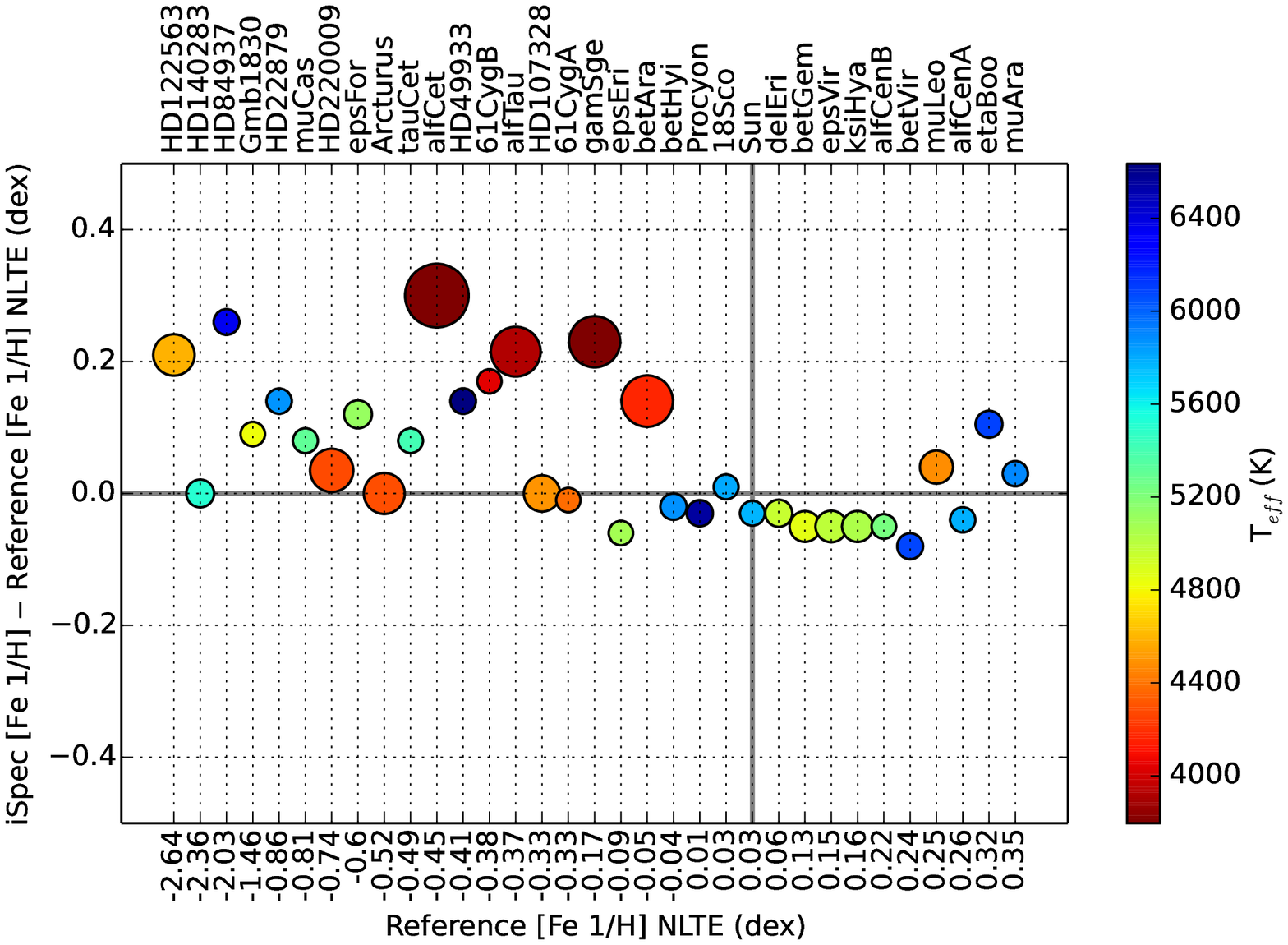}
        \par
    \end{centering}
    \caption{Differences in neutral iron abundances between the reference (the Gaia FGK benchmark stars) and the derived value by iSpec, synthetic spectral fitting method (left) and equivalent width method (right) when using the atmospheric parameters found (top) and the reference values (bottom). Stars are sorted by metallicity; the color represents the temperature, and the size is linked to the surface gravity.}
    \label{fig:bs_FeH}
\end{figure*}

The results are presented in Table \ref{tab:iron_abundances}, where the derived iron abundances from both methods are compared with the LTE and NLTE reference values. Thanks to the differential approach, part of the NLTE effects seems to cancel out (specially for solar-type stars), and the derived results agree better with these reference values. A trend depending on the reference iron abundances is observed in the top plots from Fig. \ref{fig:bs_FeH}. Both methods overestimate iron abundances for metal-poor stars.

Since errors in the atmospheric parameters propagate into the abundance determination, we repeated the analysis by fixing the effective temperature, surface gravity, and metallicity to the reference values (bottom plots in Fig. \ref{fig:bs_FeH}). The trend is still present although it is less steep. The remaining effect could still be caused by NTLE effects, which become more important at low [Fe/H], and the normalization process where the continuum placement is strongly influenced by the stellar type and metallicity (e.g., the most discrepant stars are those with the coolest and lowest surface gravity ones). For instance, if we re-normalize the spectra by forcing the continuum to be placed 1\% lower, the overall metallicity decreases by $\sim$0.07 dex when analyzed with the synthetic spectral fitting pipeline.

\begin{table}
    \caption{Average differences and standard deviation between the iron abundances derived from the two methods and the reference values (LTE and NLTE).}
    \label{tab:iron_abundances}
    \begin{center}
        \tabcolsep=0.08cm
        \begin{tabular}{l|c|c c|c c}
        \hline
               &      & \multicolumn{2}{c|}{\textbf{$\Delta$[X/H]$_{\mathrm{LTE}}$}} & 
                        \multicolumn{2}{c}{\textbf{$\Delta$[X/H]$_{\mathrm{NLTE}}$}} \\
        \textbf{} & \textbf{Element} & $\mu$ & $\sigma$ & $\mu$ & $\sigma$  \\
        \hline

\multirow{2}{*}{Synthetic spectral fitting} &   Fe 1    &   0.04    &   0.14    &   0.02    &   0.13    \\
    &   Fe 2    &   0.09    &   0.18    &   0.07    &   0.16    \\
\hline
\multirow{2}{*}{Equivalent width}   &   Fe 1    &   0.09    &   0.19    &   0.06    &   0.19    \\
    &   Fe 2    &   0.12    &   0.27    &   0.11    &   0.17    \\

        \hline                                                  
        \end{tabular}
    \end{center}
\end{table}

\section{Additional validations}\label{s:analysis}

\subsection{Model atmosphere interpolation}

The model atmosphere interpolation processes were tested by removing a list of existing pre-computed models from the original MARCS grid and forcing their interpolation. This way, we compared the interpolated version calculated by iSpec with the original pre-computed models. 

The models represent different stellar types (i.e., giants and dwarfs; metal-rich and poor) and they were selected not to be actually surrounded by any gap (i.e., missing model), as shown in Fig.~\ref{fig:atm_layer0}. The test was performed for 42 model atmospheres that were the results from different combinations of the following parameters:

\begin{enumerate}
    \item Effective temperature: 4750 and 3400 K
    \item Surface gravity: 1.00, 2.00, 2.50, 3.00, 4.00 and 4.50
    \item Metallicity: 1.00, 0.00, $-$1.00, $-$2.00, $-$3.00 and $-$4.00.
\end{enumerate}

\begin{figure}
    \begin{centering}
        \includegraphics[width=\linewidth, trim = 1mm 1mm 1mm 1mm, clip]{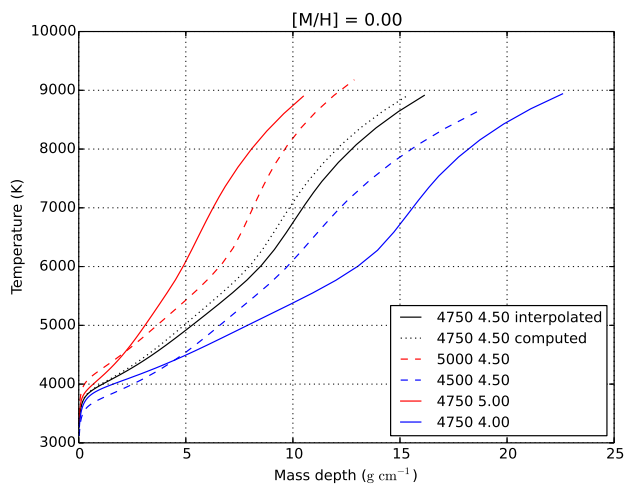}
        \par
    \end{centering}
    \caption{Temperature (K) as a function of mass depth (g cm$^{-1}$) for a computed (dotted black) and an interpolated atmosphere (solid black) from upper and lower effective temperature models (dashed red and blue) and upper and lower surface gravity models (solid red and blue).}
    \label{fig:atm_interpolated_example}
\end{figure}

The results show that the interpolated model atmosphere reproduces the overall shape of the pre-computed model (Fig. \ref{fig:atm_interpolated_example}) with a small shift.

\begin{figure}
    \begin{centering}
        \includegraphics[width=\linewidth, trim = 1mm 75mm 1mm 1mm, clip]{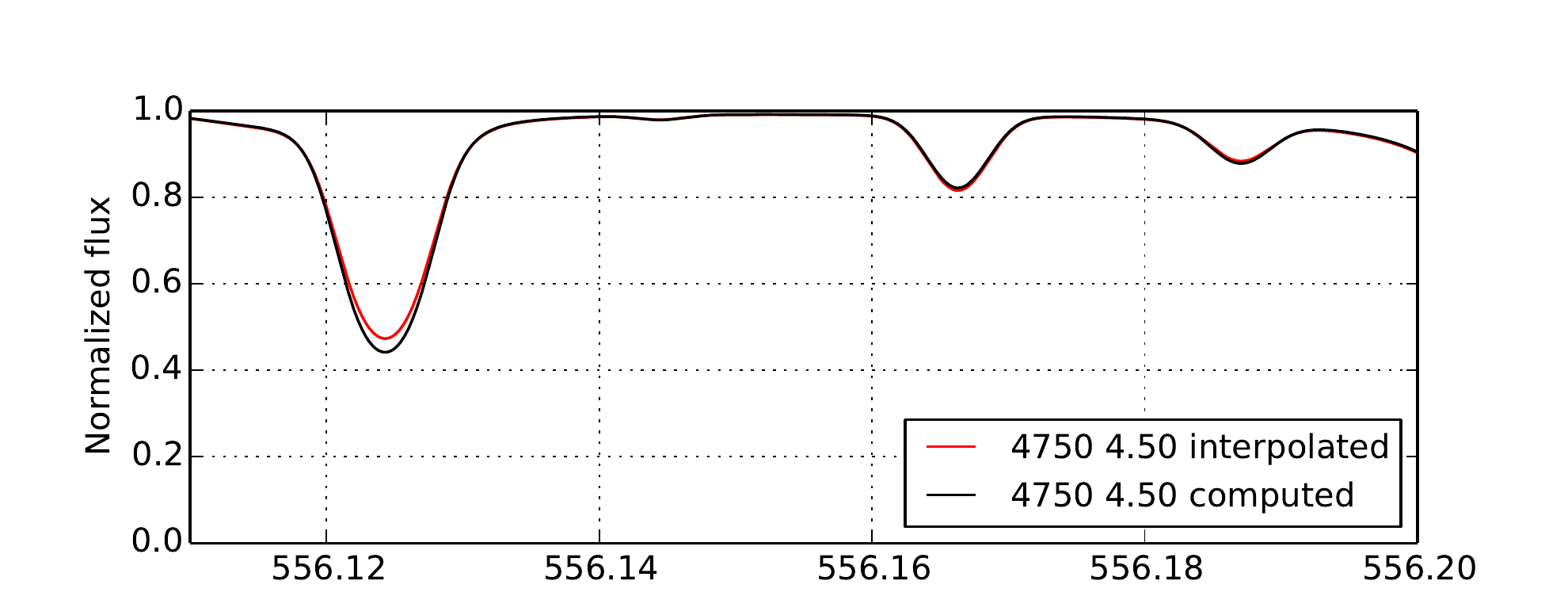}
        \par
    \end{centering}
    \caption{Synthetic spectra using a computed (black) and interpolated (red) model atmosphere.}
    \label{fig:interpolated_synth}
\end{figure}

To understand the impact of these shifts, we generated pairs of synthetic spectra (with the interpolated and the pre-computed model) and compared their fluxes (ignoring regions near to the continuum). We found that the average flux difference is $-$0.22\%$\pm$2.12\%. The spectra only differ in the central depth of the absorption lines (see Fig. \ref{fig:interpolated_synth}). Additionally, we measured the equivalent width of a group of absorption lines and determined that the average difference is 0.10\%$\pm$6.32\%. As a reference, \cite{2014A&A...566A..98B} estimated that the abundance analysis based on equivalent width methods shows a very small variation on the order of $\pm$0.007 dex in metallicity when equivalent widths are changed by 1\% (based on the analysis of a solar spectrum).

\subsection{Minimization process}

\subsubsection{Iterations}\label{s:iterations}

The minimization process consists of several iterations where the free parameters are modified by a single amount to predict the next jump in the parameter space (see Sect.~\ref{s:data_treatment}). In Table \ref{tab:results_per_iteration}, we show the evolution of the differences compared with the reference values at each iteration and the cumulative number of spectra that have already converged toward a solution for the best configurations identified in Sections \ref{s:analysis_SSF} and \ref{s:analysis_ew_method}.

The synthetic spectral fitting technique seems to present good average results around the fifth to sixth iteration. Remarkably, the dispersion in effective temperature and surface gravity deteriorates from iteration six to ten in favor of a small improvement in the metallicity dispersion. The equivalent width method presents a more gradual evolution and the results do not stabilize until the nineth or tenth iteration.

When the number of iteration is limited to a maximum of 10, the average computation time for the synthetic spectral fitting technique is 36 minutes per spectrum, while for the equivalent width method it is about 19 minutes per spectrum\footnote{Time estimated by using a computer with an Intel Xeon CPU at 3.07GHz (model X5675).} (see also Sect. \ref{s:method_comparison}).

\begin{table*}[ht!]
    \caption{Difference between the parameters derived from the two methods and the reference values on each iteration of the least-squares minimization process. The number of converged spectra per iteration is also included (78 in total).}
    \label{tab:results_per_iteration}
    \begin{center}
        \begin{tabular}{r|c c c c c c c|c c c c c c c}
        \hline
         & \multicolumn{7}{c}{\textbf{Synthetic spectral fitting}} & \multicolumn{7}{|c}{\textbf{Equivalent width}}\\
                      & \multicolumn{2}{c}{\textbf{$\Delta$T$_{\mathrm{eff}}$}} & 
                        \multicolumn{2}{c}{\textbf{$\Delta$log(g)}} &
                        \multicolumn{2}{c}{\textbf{$\Delta$[M/H]}} & 
                        \textbf{Converged} & 
                        \multicolumn{2}{|c}{\textbf{$\Delta$T$_{\mathrm{eff}}$}} & 
                        \multicolumn{2}{c}{\textbf{$\Delta$log(g)}} & 
                        \multicolumn{2}{c}{\textbf{$\Delta$[M/H]}} &
                        \textbf{Converged} \\
        \textbf{Iteration} & $\mu$ & $\sigma$ & $\mu$ & $\sigma$ & $\mu$ & $\sigma$ & & $\mu$ & $\sigma$ & $\mu$ & $\sigma$ & $\mu$ & $\sigma$ & \\
        \hline

1   &   36  &   530 &   0.30    &   0.87    &   0.03    &   0.36    &   0    (  0\%  )   &   $-$15 &   560 &   0.22    &   1.01    &   0.20    &   0.43    &   0    (  0\%  )   \\
2   &   $-$52 &   177 &   0.12    &   0.50    &   0.05    &   0.23    &   0    (  0\%  )   &   101 &   283 &   0.26    &   0.48    &   0.13    &   0.28    &   21   (  27\% )   \\
3   &   $-$12 &   117 &   0.00    &   0.32    &   0.03    &   0.16    &   3    (  4\%  )   &   146 &   271 &   0.24    &   0.47    &   0.14    &   0.26    &   38   (  49\% )   \\
4   &   $-$12 &   124 &   $-$0.08   &   0.19    &   0.02    &   0.15    &   14   (  18\% )   &   139 &   267 &   0.23    &   0.47    &   0.13    &   0.26    &   63   (  81\% )   \\
5   &   $-$14 &   121 &   $-$0.08   &   0.18    &   0.02    &   0.15    &   41   (  53\% )   &   140 &   266 &   0.23    &   0.46    &   0.12    &   0.26    &   67   (  86\% )   \\
6   &   $-$15 &   120 &   $-$0.08   &   0.17    &   0.02    &   0.15    &   60   (  77\% )   &   137 &   259 &   0.22    &   0.45    &   0.12    &   0.26    &   70   (  90\% )   \\
7   &   $-$16 &   119 &   $-$0.08   &   0.17    &   0.02    &   0.15    &   63   (  81\% )   &   133 &   251 &   0.21    &   0.45    &   0.12    &   0.26    &   71   (  91\% )   \\
8   &   $-$19 &   120 &   $-$0.09   &   0.17    &   0.02    &   0.15    &   69   (  88\% )   &   134 &   244 &   0.20    &   0.45    &   0.12    &   0.26    &   73   (  94\% )   \\
9   &   $-$22 &   122 &   $-$0.10   &   0.19    &   0.02    &   0.15    &   70   (  90\% )   &   136 &   241 &   0.20    &   0.44    &   0.12    &   0.27    &   76   (  97\% )   \\
10  &   $-$24 &   124 &   $-$0.11   &   0.21    &   0.01    &   0.14    &   78   (  100\%    )   &   135 &   241 &   0.20    &   0.44    &   0.12    &   0.27    &   78   (  100\%    )   \\

        \hline                                                  
        \end{tabular}
    \end{center}
\end{table*}

\subsubsection{Correlations}

Previous studies \citep[e.g.,][]{2012ApJ...757..161T} found strong correlations between effective temperature, metallicity, and surface gravities when simultaneously solving for all three quantities with synthetic spectral fitting techniques, while the equivalent width methods did not show such a strong correlation.

\begin{figure*}
    \begin{centering}
        \includegraphics[width=9cm, trim = 1mm 1mm 10mm 5mm, clip]{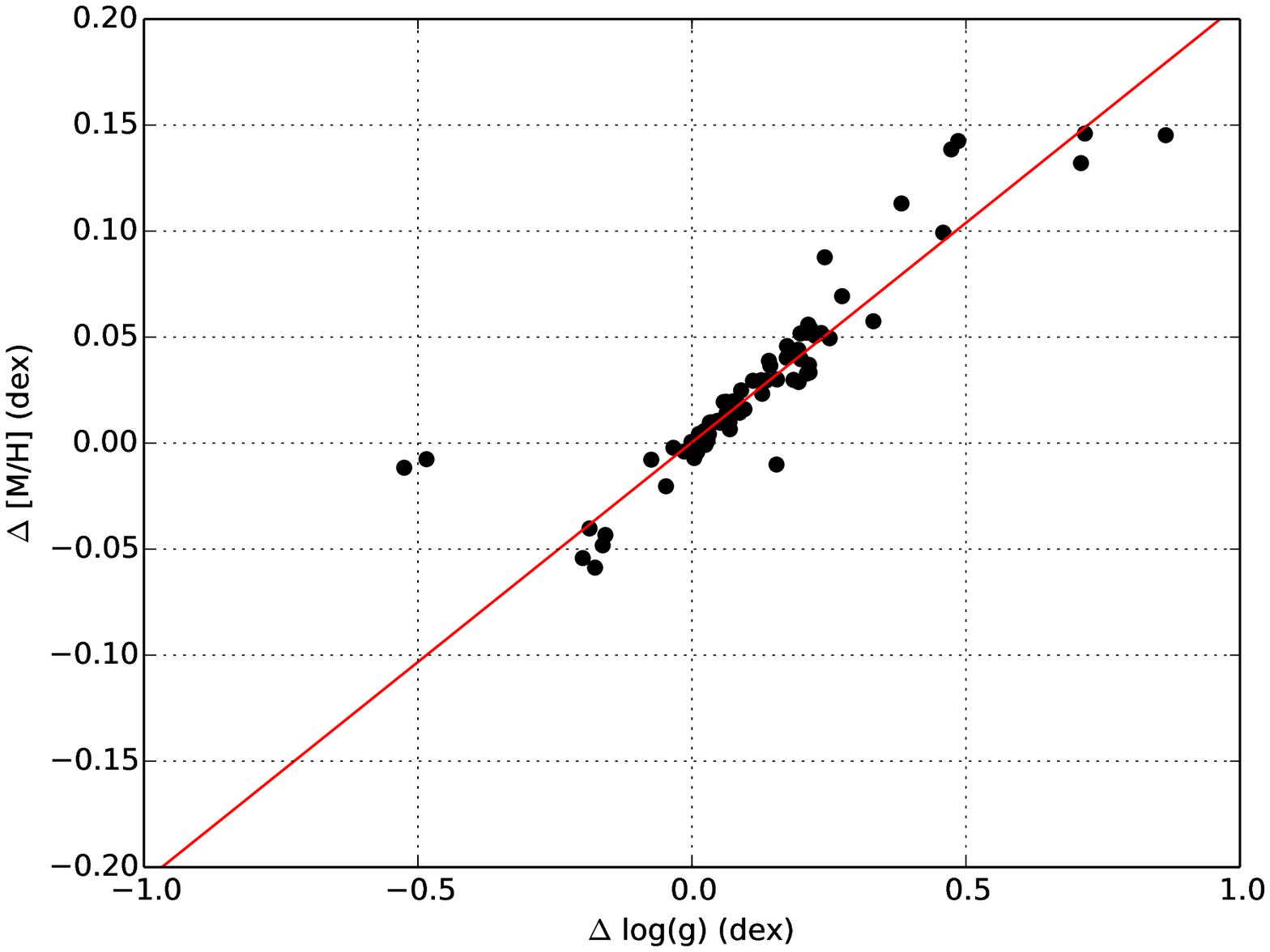}
        \includegraphics[width=9cm, trim = 1mm 1mm 10mm 5mm, clip]{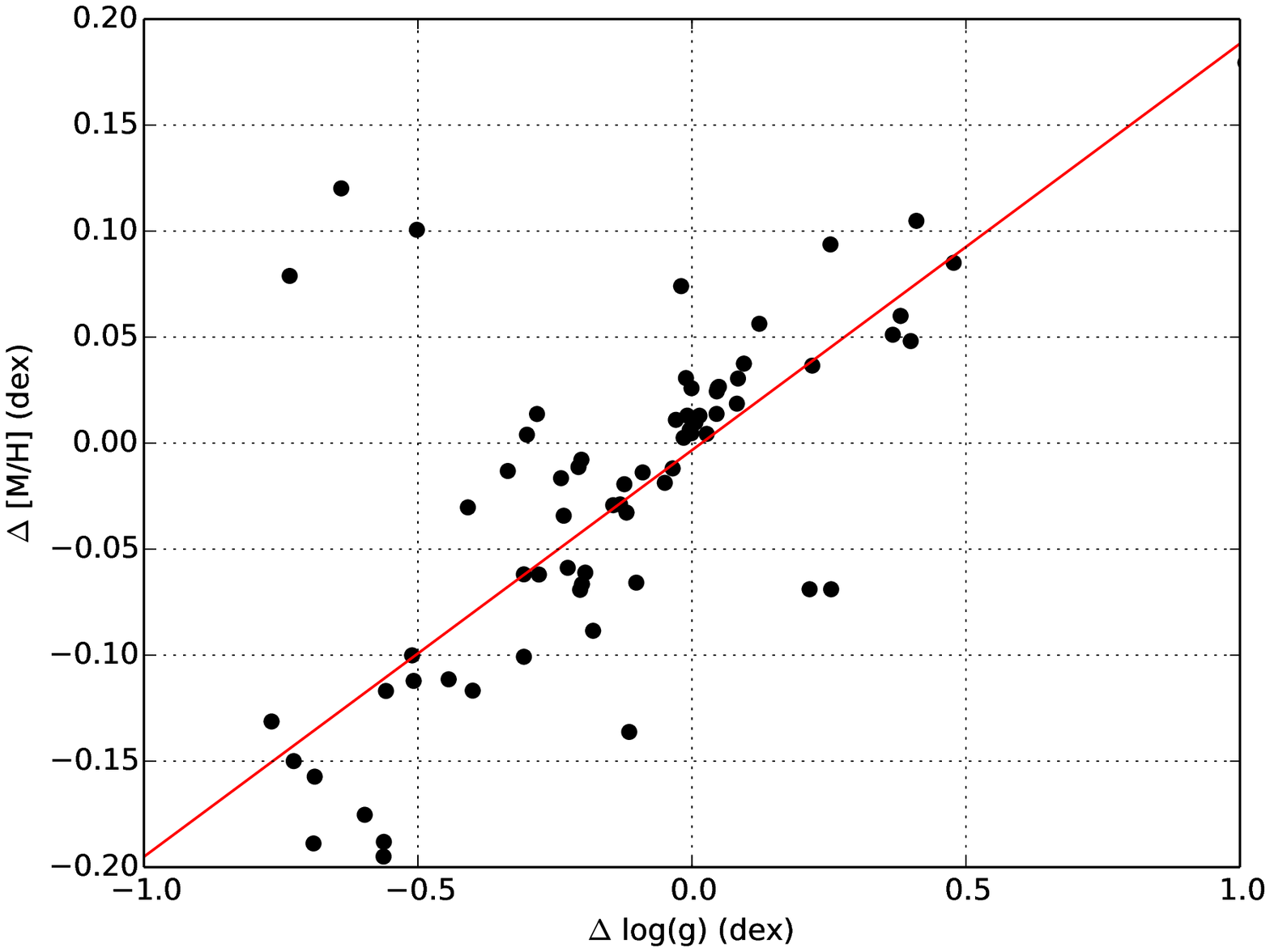}
        \par
        \includegraphics[width=9cm, trim = 1mm 1mm 10mm 5mm, clip]{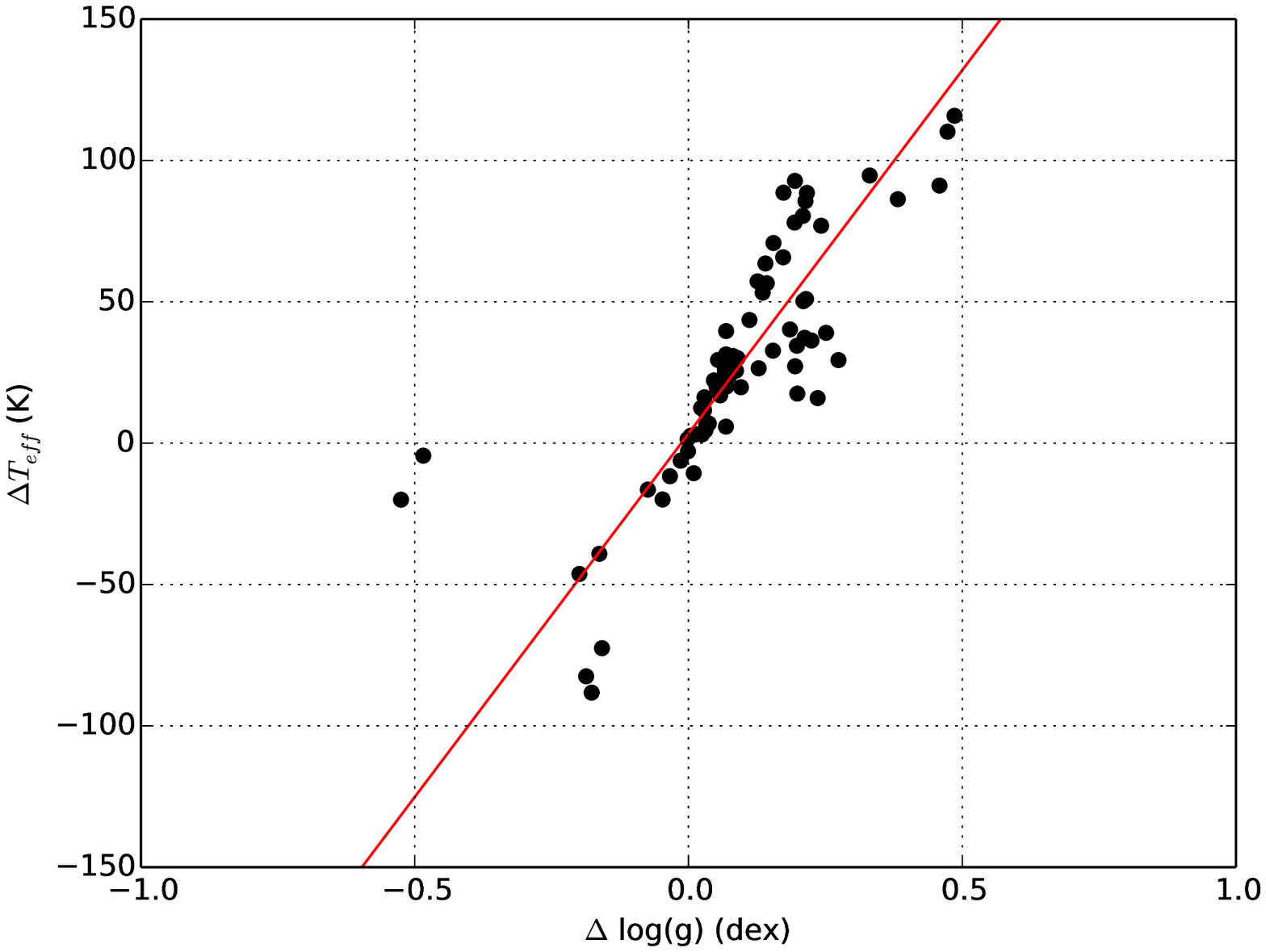}
        \includegraphics[width=9cm, trim = 1mm 1mm 10mm 5mm, clip]{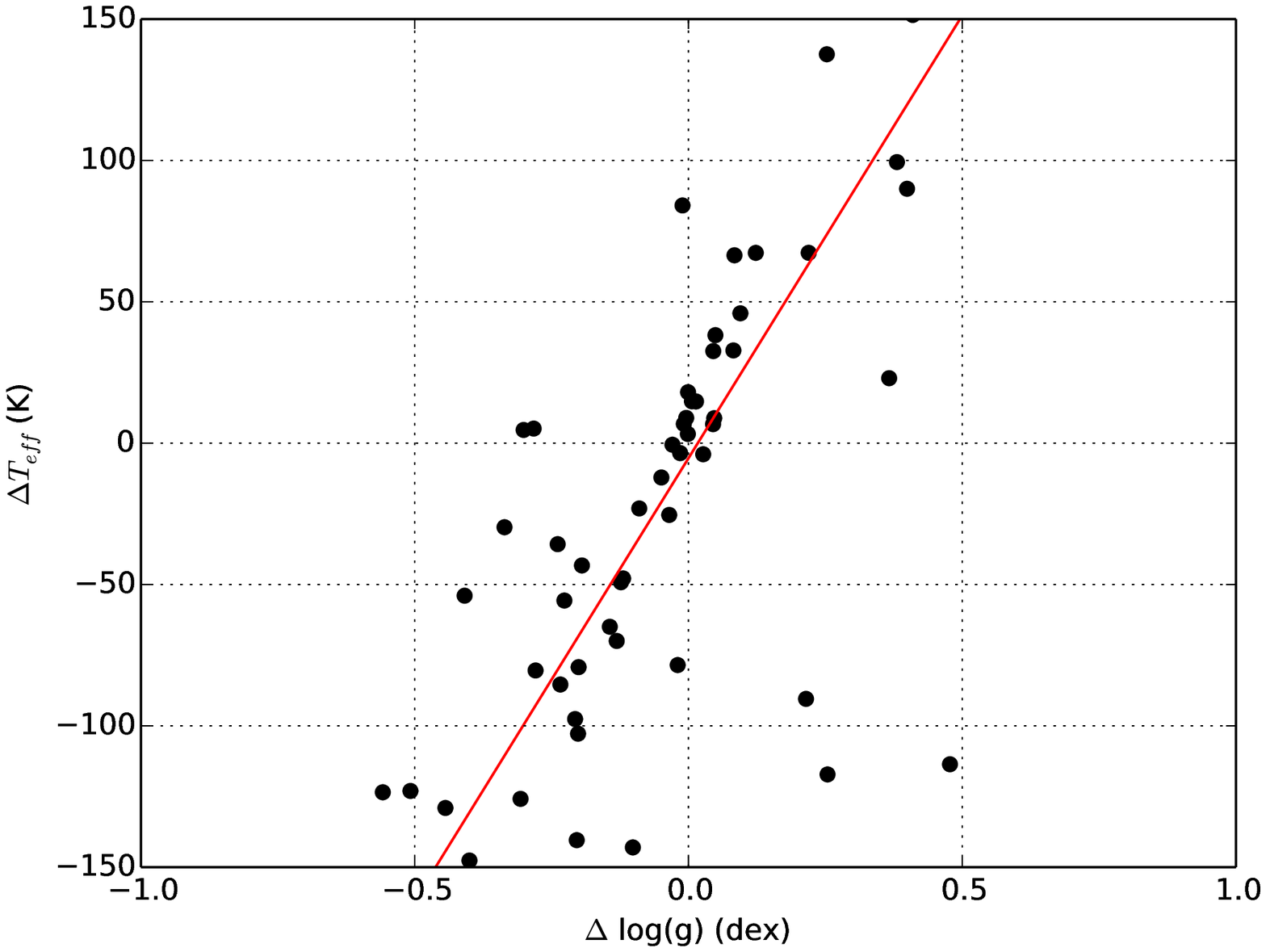}
        \par
    \end{centering}
    \caption{
Impact on the effective temperatures and metallicity derived from synthetic spectral fitting (left) and equivalent width method (right) when fixing the surface gravity to the reference value. A linear model was fitted (red line) and the differences correspond to the constrained minus unconstrained values.}
    \label{fig:correlations}
\end{figure*}

We repeated the same analysis with our Gaia FGK benchmark stars. We rederived the effective temperatures and metallicities with surface gravities fixed to the reference value and compared them with the unconstrained results (Fig.~\ref{fig:correlations}).

The parameters are obviously correlated (see Pearson correlation coefficients in Table \ref{tab:correlations}), but we see no strong differences that would depend on the method used for the analysis. However, the correlation is tighter for the spectral fitting technique. A change in surface gravity of 0.50 dex implies a difference in effective temperature of 129 and 157 K for the synthetic spectral fitting technique and the equivalent width method, respectively. For the metallicity, the impact is 0.10 dex for both methods.

\begin{table}
    \caption{Linear model parameters ($y=mx+c$) fitted using the differences on the effective temperatures/metallicity and free or fixed surface gravity shown in Fig.~\ref{fig:correlations}. The Pearson correlation coefficients are also included.}
    \label{tab:correlations}
    \begin{centering}
        \begin{center}
            \begin{tabular}{l|c| c c c}
                \multicolumn{2}{c|}{} & \textbf{$m$} & \textbf{$c$} & Pearson\\
            \hline
            \multirow{2}{*}{Synthetic spectral fitting}   &   T$_{\mathrm{eff}}$    &   257 &   3   &     0.88    \\
              &   [M/H]   &   0.21    &   0.00    &      0.90    \\
            \hline
            \multirow{2}{*}{EW}  &   T$_{\mathrm{eff}}$    &   313 &   $-$5  &    0.72    \\
              &   [M/H]   &   0.19    &   0.00    &    0.74    \\
            \end{tabular}
        \end{center}

        \par
    \end{centering}
\end{table}

It is worth noting that our analysis covers a broader parameter range: the Gaia FGK benchmark stars cover from 6600 to 3500 K in effective temperature, from 4.6 to 0.6 dex in surface gravity, and from 0.3 to $-$2.7 dex in metallicity, while \cite{2012ApJ...757..161T} analyzed stars between 6750 and 4800 K in effective temperature, 4.80 and 3.60 dex in surface gravity, and 0.5 and $-$0.4 dex in metallicity.

\subsection{Signal-to-noise ratios and spectral resolutions}

The Gaia FGK benchmark stars library consists of high-resolution and high signal-to-noise ratio spectra, but we also tested the limits of iSpec by analyzing spectra of lower quality. To do this, we generated and analyzed 34 synthetic spectra using the reference parameters of the Gaia FGK benchmark stars with different resolutions, and we added several levels of Poisson noise.

Our implementation of the synthetic spectral fitting technique seems to be more robust to noise than the equivalent width method (Table \ref{tab:different_snr_levels}). The former derives atmospheric parameters very similar to the reference values with a signal-to-noise ratio of 25, while the latter constantly deviated, even with the highest signal levels. One possible reason is that the equivalent width method simplifies the problem by considering only the area of the absorption lines and not the whole profile. Measuring that area is quite sensitive to noise and errors in continuum placement. Then, the least-squares algorithm does not minimize the difference of N fluxes (as does the synthetic spectral fitting technique), but only three values: the two slopes to validate the ionization and excitation equilibrium, and the difference between neutral and ionized iron abundance. This information loss may increase degeneracies between atmospheric parameters. Additionally, the tested implementation of the equivalent width method is strongly affected by the starting point in the parameter space. A incorrect initial estimate can lead to a poor solution. This also explains why the results for the spectra with a signal-to-noise ratio of 50 are poorer than those with a ratio of 40, for instance.

Regarding the spectral resolution, a lower resolution implies a higher number of blended lines. This has a higher negative impact on the equivalent width method, since abundances will be overestimated, while synthetic spectral fitting can reproduce and better match the blends even with resolutions as low as 7500 (Table \ref{tab:different_resolutions}).

\begin{table*}[ht!]
    \caption{Difference between the parameters derived from the two methods (using degraded synthetic spectra instead of observations) and the reference values for different levels of signal-to-noise ratio (S/N). The synthetic spectra had a resolution of 70,000.}
    \label{tab:different_snr_levels}
    \begin{center}
        \begin{tabular}{r|c c c c c c|c c c c c c}
        \hline
         & \multicolumn{6}{c}{\textbf{Synthetic spectral fitting}} & \multicolumn{6}{|c}{\textbf{Equivalent width}}\\
                      & \multicolumn{2}{c}{\textbf{$\Delta$T$_{\mathrm{eff}}$}} & 
                        \multicolumn{2}{c}{\textbf{$\Delta$log(g)}} &
                        \multicolumn{2}{c}{\textbf{$\Delta$[M/H]}} & 
                        \multicolumn{2}{|c}{\textbf{$\Delta$T$_{\mathrm{eff}}$}} & 
                        \multicolumn{2}{c}{\textbf{$\Delta$log(g)}} & 
                        \multicolumn{2}{c}{\textbf{$\Delta$[M/H]}} \\
        \textbf{S/N} & $\mu$ & $\sigma$ & $\mu$ & $\sigma$ & $\mu$ & $\sigma$ & $\mu$ & $\sigma$ & $\mu$ & $\sigma$ & $\mu$ & $\sigma$ \\
        \hline
        5   &   129 &   129 &   0.15    &   0.22    &   0.23    &   0.22    &   510 &   427 &   1.16    &   0.96    &   0.28    &   0.58    \\
        10  &   39  &   44  &   0.08    &   0.08    &   0.04    &   0.04    &   285 &   486 &   0.71    &   0.74    &   0.23    &   0.55    \\
        20  &   13  &   16  &   0.02    &   0.02    &   0.00    &   0.02    &   219 &   365 &   0.50    &   0.72    &   0.15    &   0.40    \\
        25  &   7   &   12  &   0.01    &   0.03    &   0.00    &   0.01    &   177 &   392 &   0.50    &   0.63    &   0.14    &   0.38    \\
        30  &   6   &   10  &   0.01    &   0.03    &   0.00    &   0.01    &   133 &   272 &   0.46    &   0.64    &   0.06    &   0.27    \\
        40  &   0   &   6   &   0.00    &   0.02    &   $-$0.01   &   0.01    &   149 &   293 &   0.39    &   0.61    &   0.02    &   0.21    \\
        50  &   3   &   5   &   0.00    &   0.02    &   $-$0.01   &   0.01    &   111 &   346 &   0.30    &   0.95    &   $-$0.01   &   0.26    \\
        100 &   2   &   5   &   0.00    &   0.01    &   0.00    &   0.01    &   155 &   261 &   0.37    &   0.59    &   $-$0.01   &   0.22    \\
        \hline                                                  
        \end{tabular}
    \end{center}
\end{table*}

\begin{table*}[ht!]
    \caption{Difference between the parameters derived from the two methods (using synthetic spectra) and the reference values for different resolution levels. The synthetic spectra had a signal-to-noise ratio of 100. From lower to higher resolution, they approximately correspond to the surveys SDSS DR7 (R=1800-2200), Rave, Gaia RVS, GES Giraffe HR21, APOGEE, HERMES/GALAH, GES UVES and the standard resolution of the Gaia FGK benchmark stars library.}
    \label{tab:different_resolutions}
    \begin{center}
        \begin{tabular}{r|c c c c c c|c c c c c c}
        \hline
         & \multicolumn{6}{c}{\textbf{Synthetic spectral fitting}} & \multicolumn{6}{|c}{\textbf{Equivalent width}}\\
                      & \multicolumn{2}{c}{\textbf{$\Delta$T$_{\mathrm{eff}}$}} & 
                        \multicolumn{2}{c}{\textbf{$\Delta$log(g)}} &
                        \multicolumn{2}{c}{\textbf{$\Delta$[M/H]}} & 
                        \multicolumn{2}{|c}{\textbf{$\Delta$T$_{\mathrm{eff}}$}} & 
                        \multicolumn{2}{c}{\textbf{$\Delta$log(g)}} & 
                        \multicolumn{2}{c}{\textbf{$\Delta$[M/H]}} \\
        \textbf{Resolution} & $\mu$ & $\sigma$ & $\mu$ & $\sigma$ & $\mu$ & $\sigma$ & $\mu$ & $\sigma$ & $\mu$ & $\sigma$ & $\mu$ & $\sigma$ \\
        \hline
        2000    &   56  &   192 &   0.14    &   0.67    &   $-$0.03   &   0.17    &   530 &   923 &   0.10    &   1.39    &   0.25    &   1.67    \\
        7500    &   2   &   13  &   0.00    &   0.03    &   0.01    &   0.01    &   629 &   874 &   $-$0.16   &   1.53    &   0.12    &   1.97    \\
        11500   &   4   &   6   &   0.00    &   0.02    &   0.01    &   0.01    &   769 &   574 &   0.69    &   1.42    &   0.53    &   0.39    \\
        16200   &   1   &   5   &   0.00    &   0.01    &   0.01    &   0.01    &   450 &   600 &   0.88    &   1.24    &   0.28    &   0.32    \\
        20000   &   1   &   3   &   0.00    &   0.01    &   0.00    &   0.00    &   324 &   440 &   0.82    &   1.10    &   0.25    &   0.34    \\
        28000   &   1   &   4   &   0.00    &   0.01    &   0.00    &   0.00    &   267 &   292 &   0.69    &   0.87    &   0.15    &   0.21    \\
        47000   &   1   &   4   &   0.00    &   0.01    &   0.00    &   0.01    &   218 &   217 &   0.47    &   0.57    &   0.06    &   0.23    \\
        70000   &   2   &   5   &   0.00    &   0.01    &   0.00    &   0.01    &   155 &   261 &   0.37    &   0.59    &   $-$0.01   &   0.22    \\
        \hline                                                  
        \end{tabular}
    \end{center}
\end{table*}

\section{Conclusions}\label{s:conclusions}

iSpec is an integrated spectroscopic software framework with the necessary functions for determining of atmospheric parameters (i.e., effective temperature, surface gravity, metallicity) and individual chemical abundances. It relies on the widely known code SPECTRUM developed by R. O. Gray for spectral synthesis and derivation of abundances from equivalent widths. 

We developed two different pipelines based on the synthetic spectral fitting technique and the equivalent width method by using iSpec. The high-resolution and high signal-to-noise spectra from the Gaia FGK benchmark star library were used to assess the pipelines. We showed the following:

\begin{enumerate}
    \item The pipeline based on the synthetic spectral fitting technique provides more accurate and precise results.
    \item The derived effective temperature, surface gravity, and metallicity parameters are correlated at a similar degree, independently of the technique used on each pipeline.
    \item The pipeline based on synthetic spectral fitting technique is more effective with lower resolutions (i.e., 7500) and lower signal-to-noise ratios (i.e., 25) than the pipeline based on equivalent widths.
\end{enumerate}

Additionally, we showed how the Gaia FGK Benchmark Star library can be used to assess and optimize spectroscopic pipelines. Taking advantage of this set of stars to verify and improve pipelines for spectroscopic analysis can lead to more reliable and comparable results.

\begin{acknowledgements}
    We thank L. Chemin, N. Brouillet, D. Mahdi, and T. Jacq for their feedback as iSpec users, and R. O. Gray for his code SPECTRUM. We also thank B. Smalley for the valuable comments as referee.
    This work was partially supported by the Gaia Research for European Astronomy Training (GREAT-ITN) Marie Curie network, funded through the European Union Seventh Framework Programme [FP7/2007-2013] under grant agreement n. 264895.
    UH acknowledges support from the Swedish National Space Board (Rymdstyrelsen).
    All the software used in the data analysis were provided by the Open Source community.
\end{acknowledgements}


\bibliographystyle{bibtex/aa} 
\bibliography{References} 

\end{document}